\documentclass[11pt,a4paper]{article}
\usepackage{amssymb}
\usepackage{amsmath}
\usepackage{amsthm}
\usepackage[utf8]{inputenc}
\usepackage{physics}
\usepackage{enumitem}
\usepackage{mathtools}
\usepackage{bbold}
\usepackage{cite}
\usepackage{subcaption}

\usepackage{hyperref}
\hypersetup{
    colorlinks,
    citecolor=blue,
    filecolor=red!75!black,
    linkcolor=red!75!black,
    urlcolor=blue
}

\numberwithin{equation}{section}

\usepackage{tikz-feynman}
\tikzfeynmanset{compat=1.1.0}

\usepackage{appendix}

\usepackage{authblk}

\theoremstyle{definition}

\newcommand{\la}{\lambda}
\newcommand{\lat}{\Tilde{\lambda}}
\newcommand{\rhot}{\Tilde{\rho}}

\newcommand{\M}{\mathcal{M}}
\newcommand{\A}{\mathcal{A}}

\newcommand{\agl}[2]{\langle #1 #2 \rangle}
\newcommand{\sqr}[2]{\lbrack #1 #2 \rbrack}

\DeclareMathAlphabet{\mathmybb}{U}{bbold}{m}{n}

\begin{document}

\begin{center}

\begin{flushright}
{\small Saclay-t24/014}\\
\end{flushright}

\vspace{2cm}
{\Large \bf Continuous-Spin Particles, On Shell}

\vspace{1.4cm}
{Brando Bellazzini,  Stefano De Angelis, and Marcello Romano}

 \vspace*{.5cm} 
\begin{footnotesize}
\begin{it}
 Universit\'e Paris-Saclay, CNRS, CEA, Institut de Physique Th\'eorique, 91191, Gif-sur-Yvette, France.  \\
\end{it}
\end{footnotesize}

\vspace*{10mm}

\begin{abstract}\noindent\normalsize
We study on-shell scattering amplitudes for continuous-spin particles (CSPs). Poincar\'e  invariance, little-group $ISO(2)$ covariance, analyticity, and on-shell factorisation (unitarity) impose stringent conditions on these amplitudes. We solve them by realizing a non-trivial representation for all little-group generators on the space of functions of bi-spinors.
The three-point amplitudes are uniquely determined by matching their high-energy limit to that of definite-helicity (ordinary) massless particles. Four-point amplitudes are then bootstrapped using consistency conditions, allowing us to analyze the theory in a very transparent way, without relying on any off-shell Lagrangian formulation. We present several examples that highlight the main features of the resulting scattering amplitudes. 
We discuss CSP's amplitudes as a new infrared deformation of ordinary massless amplitudes, which is controlled by the scale of the Pauli-Lubanski vector squared, as opposed to the familiar mass deformation.  
Finally, we explore under which conditions it is possible to relax some assumptions, such as strict on-shell factorisation, analyticity, or others. In particular,  we also investigate how continuous-spin particles may couple to gravity and electromagnetism, in a loose version of $S$-matrix principles.

\end{abstract}

\end{center}

\newpage

\renewcommand{\baselinestretch}{0.9}\normalsize
{ \hypersetup{hidelinks} 
 \tableofcontents }
\renewcommand{\baselinestretch}{1}\normalsize

\newpage

\section{Introduction}
\label{Sec:Intro}

A promising strategy for understanding fundamental interactions relies on studying the consistency conditions of scattering amplitudes. This approach seeks to determine which theories adhere to foundational principles that are stringent enough to filter out most candidates, yet permissive enough to accommodate a few (or unique) viable scattering amplitudes. This line of reasoning has proven very fruitful in the context of relativistic quantum scattering in Minkowski space, particularly for massless particles with spin.

For instance, it is well-established that any theory describing massless spin-1 particles must be a gauge theory while the theory of a massless spin-2 particle is unique in its infrared limit, corresponding to general relativity (GR) \cite{Weinberg:1965nx,Weinberg:1964ew}. Similarly, the theory of massless spin-3/2 particles must be a supergravity theory \cite{Grisaru:1977kk}, while massless particles with spins greater than 2  have a trivial $S$-matrix \cite{Porrati:2008rm}, a result extending the Weinberg-Witten constraints on massless particles \cite{Weinberg:1980kq}.  
%
%
In recent years, positivity bounds 
and $S$-matrix bootstrap constraints have further shaped the landscape of consistent theories, extending some of the previous results to massive spinning particles, see \textit{e.g.} \cite{Bellazzini:2023nqj,Bellazzini:2019bzh,Bertucci:2024qzt,Davighi:2021osh}.

In a sense, these constraints can be viewed as modern selection rules derived directly from the fundamental principles of quantum mechanics and relativity.

However, amidst this well-explored landscape, there exists a class of massless particles that has received considerably less attention: continuous-spin particles (CSPs). CSPs are massless irreducible unitary representations (irreps) of the Poincar\'e  group, classified long ago by Wigner \cite{Wigner:1939cj}. As the associated little group (LG) is non-compact and faithfully represented, a CSP carries infinitely many degrees of freedom even at fixed momentum.

The theory of CSPs has been found incompatible with the stringent axioms of local field theory, see \textit{e.g.} \cite{Yngvason:1970fy, Iverson:1971hq, Abbott:1976yz}. However, it was realized years later that the assumptions could be significantly relaxed, as the axiomatic approach would otherwise reject ordinary gauge and gravity theories. A free Lagrangian gauge-theory formulation emerged \cite{Schuster:2013pxj, Schuster:2013vpr, Schuster:2013pta, Schuster:2014hca} along with some encouraging results extracted from soft limits of the scattering amplitudes. This progress triggered further investigations, such as \footnote{For a nearly comprehensive reference list along this direction, see the references in reference~\cite{Schuster:2023jgc}.}  \cite{Rivelles:2014fsa,Rivelles:2016rwo,Metsaev:2016lhs,Metsaev:2017cuz,Buchbinder:2018soq,Buchbinder:2018yoo,Buchbinder:2020nxn,Bekaert:2017xin}, and an interacting Lagrangian theory, formulated in terms of gauge fields coupled to matter worldlines, has been recently proposed \cite{Schuster:2023jgc, Schuster:2023xqa}.

Stimulated by these intriguing advances, the purpose of this paper is twofold: (i) to understand the properties of CSPs directly through the lens of consistency conditions of on-shell scattering amplitudes; (ii) to explore which principles ---if any--- need to be relaxed for non-trivial solutions of the constraints to exist. A vital part of both points is also the derivation of explicit amplitudes that bear direct physical implications of underlying principles.

This on-shell approach is particularly well-suited to analyze CSP dynamics and scrutinize its unusual features in a very transparent way.
Indeed, it allows us to classify all three-point on-shell amplitudes consistent with LG-covariance, analyticity, and well-defined high-energy behaviour. We characterize as well the structure of all on-shell $n-$point amplitudes, and analyse their properties concerning on-shell factorisation ---extended unitarity---, analyticity, crossing symmetry, and high-energy limits. We illustrate our findings through several examples. We finally critically discuss possible ways to relax these constraints and their implications for gravity and electromagnetism.  

Interestingly, we show that CSPs produce new amplitudes that, at short distances, reproduce those of a finite number of ordinary massless particles, while at long distances the dynamics is completely changed  by the recoupling of infinitely many additional degrees of freedom. While reminiscent of a Higgs-like mechanism, the CSPs remain massless, as it is the second Casimir invariant---the Pauli-Lubanski squared operator---that is turned-on, controlling the onset of non-trivial yet massless IR dynamics.

The paper is structured as follows: Section~\ref{Sec:kinematics} introduces basic kinematic concepts and defines CSPs. In Section~\ref{sec:bootstappingGeneral}, we present the primary findings of this study,  bootstrapping amplitudes for CSPs from foundational principles. Section~\ref{sec:ExamplesGeneral} showcases non-trivial examples that highlight key features of CSPs amplitudes, while Section~\ref{sec:weakening-Assumptions} explores the relaxation of certain constraints and their resulting physical implications. Finally, we summarize our findings and explore future directions in Section~\ref{sec:discussionConclusions}.
 Our conventions are summarized in appendix~\ref{appendix:conventions}, while appendices  \ref{appendix:rhospinors}  and \ref{appendix:Fourier} contain technical details. 
 


\section{Kinematics and Poincar\'e}
\label{Sec:kinematics}

We consider a relativistic quantum theory of particles scattering in four-dimensional Minkowski space. 
We assume the dynamics is invariant under the inhomogeneous proper Lorentz group.  In particular  asymptotic in- and out-going scattering states not only cover the entire positive Hilbert space ---unitarity--- but also transform as the tensor product of single particle states which carry themselves a unitary irrep.\footnote{The case of pairwise helicity states is discussed separately in Section~\ref{sec:backgrounds}.}   Each particle $i=1,2,\ldots $ is thus associated to a pair of real numbers $(m^2_i, \mu^2_i)$  which are the values taken by the two Casimir operators  ---the mass-squared and the Pauli-Lubanski squared---
\begin{equation}
P_\mu P^\mu=m^2\ ,\qquad W_{\mu} W^{\mu}=- \mu^2 \ . 
\end{equation}
The Casimirs  are written in terms of the group generator $P^\mu$ and $J^{\mu\nu}$, and the so-called Pauli-Lubanski pseudo-vector 
\begin{equation}
\label{eq:Wdef}
 W_{\mu} \equiv \frac{1}{2}\epsilon_{\mu\nu\rho\sigma}J^{\nu\rho}P^\sigma\ ,  \qquad [W_\mu, P_\mu ]=0\ , \qquad  [W_\mu,W_\nu]=-i\epsilon_{\mu\nu\rho\sigma}W^{\rho}P^\sigma  \ . 
\end{equation}
On a subspace spanned by definite-momentum states with $ p^\mu \neq 0$, the \eqref{eq:Wdef} imply that the Pauli-Lubanski is the generator of the ``little group'' $H^{p}$ of $p^\mu$, that is the Lorentz subgroup that leaves that  definite-$p^{\mu}$ subspace invariant.

For a null momentum $p^2=0$, the $H^p\sim ISO(2)$, which is the group of isometries of 2D euclidean plane made of rotations  $\mathbb{H}$ around the direction of motion ---the helicity--- and translations $\mathbb{W}^{\pm}$ 
\begin{equation}
\label{eq:iso2algebra}
[\mathbb{H}, \mathbb{W}^{\pm} ]=\pm \mathbb{W}^{\pm}\ , \qquad  [\mathbb{W}^{+} , \mathbb{W}^{-}]=  0\ .
\end{equation}
For instance, the $ISO(2)=H^{\widehat{p}}$ generators associated to $p=\widehat{p}=(1,0,0,1)$  are $\mathbb{H}=J_3=-W_3$ and $\mathbb{W}^{\pm}=-(W_1 \pm i W_2)$. The latter are lowering and raising operators within the complexified Lorentz algebra $sl(2,\mathbb{C})\times sl(2,\mathbb{C})$.\footnote{It is convenient to work with $\mathbb{W}^{\pm}$, rather than $W_{1,2}$, but the little group is subgroup of the real $SL(2,\mathbb{C})$ Lorentz group, as in equation~\eqref{eq:little_group_generators}.}
For a massive momentum  $p^2>0$, the little group is the familiar rotation group $SU(2)$.

The little group $H^p$ is relevant because unitary irreps\footnote{Different irreps would be labelled by \textit{e.g.} the spin $j$, the helicity $h$, or $\mu^2$ as in ${D^{(j)}}^{h^\prime}_{\,\,h}$, $e^{-ih\omega}\delta_{h^\prime}^h$, and ${D^{(\mu)}}^{h^\prime}_{\,\,h}$, but we suppress this label whenever possible to avoid clutter of notation. } $D(g)$ of its universal cover,  $|p\, h\rangle \to |p\, h^\prime \rangle D^{h^\prime}_{\,\, h}(g)$, induce unitary irreps of the $SL(2,\mathbb{C})$, on generic momentum state
\begin{equation}
\label{eq:TrasformationState}
| k\, h \rangle \to  | \Lambda k\, h^\prime \rangle D^{h^\prime}_{\,\, h}(W(\Lambda,p,k)) \ ,\qquad W(\Lambda,p,k)\in H^p
\end{equation}
where the Wigner transformation $W(\Lambda,p,k)$ depends in some complicated way on the chosen reference momentum $p$, the momentum $k$ of the state, and the conventional choice of Lorentz transformation $L(p,k): p\to k$ used to define the state  $|k \, h\rangle \equiv U(L(p,k))|p\, h\rangle$. 
Different choices of $p$ or $L(p,k)$ correspond to redefining the states by a little-group transformation in $H^p$, $H^k$, or their intersection (pairwise helicity transformations). These ambiguities, which reflect the freedom of choosing the basis for each one-particle state independently, manifest themselves in the little-group covariance of scattering amplitudes with respect to each $H^{p_i}$. Then, in the following, we assume that amplitudes provide a linear space where the tensor product representation $\bigotimes_i D_i$ acts, where the sub-index $i$ distinguishes the irreps for the $i-th$ particle.

The states of definite $p$ for a single CSP can be labelled by diagonalising the helicity $\mathbb{H}$ --- the $h$-basis --- where 
\begin{equation}
\label{eq:hbasisdefinition}
\mathbb{H}| p\, h\rangle = h | p\, h\rangle \ ,\qquad \mathbb{W}^{\pm}| p \, h\rangle  = \mu | p \, h\pm 1\rangle 
\end{equation}
so that rotations multiply by a phase whereas translations mix all helicity by a Bessel function
\begin{equation}
\label{eq:hbasisDexplicit}
e^{i\alpha^{+}\mathbb{W}^{-} +i\alpha^{-}\mathbb{W}^{+}}| p\,h\rangle=\sum_{h'}\left(-\frac{\alpha_{+}}{\alpha_{-}}\right)^{\frac{h-h'}{2}}J_{h-h'}(2\sqrt{\alpha^+ \alpha^-}\mu)\, | p\,h'\rangle \,.
\end{equation}
Notice that the helicity is no longer Lorentz invariant as $\mathbb{W}_{\pm}$ can raise/lower it. It is still taking nevertheless (all) integers or (all) half-integers values. 

In the rest of the paper we find it easier to work in a basis that  diagonalises both $\mathbb{W}^{\pm}$ --- the $\theta$ basis ---
\begin{equation}
\mathbb{W}^{\pm}  |p\, \theta \rangle =\mu^{\pm}  |p\, \theta \rangle \ , \qquad \mathbb{H}|p\, \theta\rangle= i\frac{\partial}{\partial\theta}|p\,\theta\rangle\ , \qquad \mu^{\pm}\equiv \mu e^{\pm i \theta}
\end{equation}
where the state is defined by an angle $\theta$ with $2\pi$ ($4\pi$) periodicity for bosons (fermions). The $ISO(2)$-translations act multiplicatively, whereas the helicity is rotating $\theta$
\begin{equation}
\label{eq:little_group_generators}
    e^{i\alpha^{+}\mathbb{W}^{-} +i\alpha^{-}\mathbb{W}^{+}}| p\,\theta\rangle = e^{i\alpha^{+}\mu^{-} +i\alpha^{-}\mu^{+}} |p\, \theta\rangle\ ,\qquad 
    e^{-i\omega \mathbb{H}}|p\, \theta\rangle=|p\, \theta+\omega\rangle\ . 
\end{equation}
The $\theta$ and helicity-basis are connected by a simple Fourier transform:
\begin{equation}
\label{eq:changebasistoh}
    \begin{split}
        |p\,  h\rangle = \int_0^{2\pi}\!\frac{\dd \theta}{2\pi}\, e^{+ i \theta h}\, |p\, \theta\rangle = 
        \int_0^{2\pi}\!\frac{\dd \theta}{2\pi}\, e^{+ i \theta h}\sum_{\bar{h}=-\infty}^{+\infty} e^{- i \theta \bar{h}}
        |p\, \bar{h}\rangle\ .
    \end{split}
\end{equation}
%

In the following, we assume the S-matrix operator commutes with translations and Lorentz generators, $[S,P^\mu]=0=[S,J^{\mu\nu}]$. The first equality ---translation invariance--- implies momentum conservation in the form of an overall Dirac-delta  in the scattering amplitudes\begin{equation}
i (2\pi)^4\delta^4(\sum k_i) \mathcal{M}_{h_{1}\ldots h_{n}}(k_1, \ldots k_n)= \langle \mathrm{out}|S-\mathbb{I}|\mathrm{in}\rangle \ . 
\end{equation}
The second equality ---Lorentz invariance--- demands the following relation
\begin{equation}
\label{eq:lorentz}
 \mathcal{M}_{h_{1}\ldots h_{n}}(k_1, \ldots   k_n)=    \mathcal{M}_{h^\prime_{1}\ldots h^\prime_{n}}( \Lambda k_1, \ldots  \Lambda k_n) {D_{1}}^{h^\prime_1}_{\,\,\, h_1}(W(\Lambda,p,k_1))\cdots {D_{n}}^{h^\prime_n}_{\,\,\, h_n}(W(\Lambda,p,k_n))\ . 
\end{equation}
Explicitly, for an incoming CSP in $\theta$ basis the \eqref{eq:lorentz} reads  
\begin{equation}
\mathcal{M}_{\theta_1\cdots h_n}(k_1,\ldots,k_n)=\mathcal{M}_{\theta_1+\omega \cdots h_n^\prime}(\Lambda k_1,\ldots, \Lambda k_n) e^{i\alpha^+ \mu_1^- +i\alpha^- \mu_1^+} \cdots {D_{n}}^{h^\prime_n}_{\,\,\, h_n}(W(\Lambda,p,k_n))
\end{equation}
where $\omega$ and $\alpha_{\pm}$ are the ISO(2) Lie parameters associated with $W(\Lambda,p,k)$. The irrep $D$ for CSPs in $h$-basis is instead given by an helicity phase factor, like for massless particles, followed by the action of a Bessel function like in \eqref{eq:hbasisDexplicit}. 

Notice that $D_i=D^{(\mu_i)}$ for  incoming CSP  $|\mathrm{in}\rangle$, and its complex conjugate irrep $D_i=D^{*(\mu_i)}$ for a outgoing CSP $\langle \mathrm{out}|$. The distinction between incoming and outgoing irreps is in fact artificial. For instance, ordinary massless particles in the bra-vector transform like opposite helicity in the ket-vector. Likewise, CSPs in the bra-vector transform as  CSPs in the ket-vector, up to the replacement  $\theta\to\theta+\pi$, schematically $\langle p\,\,  \theta| \sim |-p\,\, \theta+\pi\rangle$ (or $\langle p\,\,  h| \to (-1)^h |-p\, \,-\!h\rangle$ in the helicity basis), see Appendix~\ref{appendix:conventions} for more details.  Analogous story for massive particles, and also for the irreps associated to internal symmetries that are as well mapped to their complex conjugate, \textit{e.g.} $q\to -q$ for $U(1)$ charges. 
The 4-momentum from outgoing to incoming is flipped  as well, treating all momenta on equal footing, the sign of the energy on real kinematics distinguishing the actual amplitude. 

Without loss of generality, we thus adopt an all-in convention where particles transform as if they were all incoming in ket-vectors $|\mathrm{in}\rangle$.

\section{Bootstrapping Amplitudes}
\label{sec:bootstappingGeneral}

\subsection{$ISO(2)$ functionals}
\label{Sec:differential}

Little group covariance and Lorentz invariance~\eqref{eq:lorentz} are the kinematic constraints on amplitudes studied in this section.
Ordinary massless and massive amplitudes solve these constraints because they are $SL(2,\mathbb{C})\times SL(2,\mathbb{C})$-invariant functions of spinors. Indeed, given a particle of null momentum, the associated spinors 
\begin{equation}
\label{eq:momentum-spinor}
|p\rangle [p|= \lambda_{\alpha} \lambda_{\dot\alpha}= p_{\alpha\dot{\alpha}}\ , \qquad p^2=0
\end{equation}
are defined only up to (complexified) little group transformations, as rescaling $\lambda(p)\to w \lambda(p)_{\alpha}$ and $\lambda_{\dot\alpha}(p)\to w^{-1}\lambda(p)_{\alpha}$ preserves the momentum. They are elements of an equivalence class $(p^\mu, p^2=0) \leftrightarrow (|p\rangle, |p])/U(1)$.  Acting with a $SL(2,\mathbb{C})\times SL(2,\mathbb{C})$ transformation on a conventionally chosen representative of the element class maps into another choice
\begin{equation}
\Lambda_{\alpha}^{\,\,\beta}\lambda_{\beta}(p)= w\lambda_{\beta}(\Lambda p)\ ,
\qquad \Lambda_{\dot\alpha}^{\,\,\dot\beta}\lambda_{\dot\beta}(p)= w^{-1}\lambda_{\dot\beta}(\Lambda p) \ .
\end{equation}
Therefore, amplitudes of the form $\mathcal{M}_{h_1\ldots h_n}(\{| i\rangle, |i]\})=\mathcal{M}\left(\langle i j \rangle, [i j]\right)$ satisfy the Lorentz invariant constraint \eqref{eq:lorentz} if 
\begin{equation}
 \mathcal{M}_{h_1\ldots h_n}(\{w_i| i\rangle, w_i^{-1}|i]\})=\left(w_{i}\right)^{2h_i}  \times \mathcal{M}_{h_1\ldots h_n}(\{| i\rangle,|i]\}) \qquad \forall i=1,\ldots n\ . 
\end{equation}
But this is precisely the $U(1)$ little-group covariant constraint, the only one left to be enforced for ordinary massless particles. It is rather simple because it amounts to count left-handed and right-handed spinors for each particle. This counting can be translated into an equivalent differential problem
\begin{equation}
\mathbb{H}_i \mathcal{M}_{h_1\ldots h_n}= h_i\mathcal{M}_{h_1\ldots h_n}\qquad \mathbb{H}_i=-\frac{1}{2}\left( |i\rangle \frac{\partial}{\partial |i\rangle}- |i] \frac{\partial}{\partial |i]}\right)\ . 
\end{equation}
The main lesson is that Lorentz invariant amplitudes are solutions of a linear differential problem defined by the little group of each particle. A completely analogous story can be told for massive spinors, see \textit{e.g.} \cite{Arkani-Hamed:2017jhn}. 

We apply now the same logic to CSPs.  We need to find three differential operators $\mathbb{H}$, $\mathbb{W}^{\pm}$ that realise the $ISO(2)$ algebra \eqref{eq:iso2algebra} on the space of functions of spinors. 

The space must be larger than the one considered above, or else $W_\mu^2=0$. Indeed, the translations operators act multiplicatively on it via $P_{\alpha\dot\alpha}=\lambda_\alpha \lambda_{\dot\alpha}$, while the little group is a Lorentz subgroup generated for each particle spinor by
$J_{\alpha\beta}=i\lambda_{(\alpha}\frac{\partial}{\partial\lambda^{\beta)}}$ and $ J_{\dot{\alpha}\dot{\beta}}=i\lambda_{(\dot{\alpha}}\frac{\partial}{\partial\lambda^{\dot{\beta})}}$. 
Using these expressions in the definition \eqref{eq:Wdef} of Pauli-Lubanski  (in spinorial form) returns vanishing Casimirs
\begin{equation}
\label{eq:PauliLubSpinorForm}
W_{\alpha\dot{\alpha}}=\frac{-i}{2}\left(J_{\alpha}^{\,\,\beta} P_{\beta \dot\alpha}- J^{\dot\beta}_{\,\,\dot\alpha}  P_{\alpha \dot\beta}\right)\longrightarrow W_{\alpha\dot{\alpha}}=P_{\alpha\dot\alpha}\mathbb{H} \longrightarrow W_{\alpha\dot\alpha}W^{\alpha\dot\alpha}=0\ ,
\end{equation}
so that a functions of $(\lambda_{\alpha},\lambda_{\dot\alpha})$-only can't describe an amplitude for a CSP. 

A large enough space of functions of spinors is instead the one that acts on spinors $\lambda_{\alpha}$ and $\lambda_{\dot\alpha}$ defined by \eqref{eq:momentum-spinor} as well as on linearly independent spinors denoted by $\rho_\alpha$ and $\rho_{\dot\alpha}$ 
\begin{equation}
\label{eq:defrho}
\langle \lambda^i \rho^i\rangle= \langle i \mathmybb{i}\rangle  \neq 0\ , \qquad [\lambda^i \rho^i]=[i \mathmybb{i}]\neq 0 \ . \end{equation}
We refer to these linear independent spinors as {\it ``black-board'' } angle (square) spinors $|\mathmybb{i}\rangle$ ($[\mathmybb{i}|$). 
They always exist because the space of spinors $\mathbb{C}^2$ is two-dimensional.\footnote{A concrete example for massless momentum pointing along $z$-direction is $\lambda_{\alpha}=\sqrt{2E}\left(\begin{array}{c}0\\ 1\end{array}\right)
$, $\rho_{\alpha}=\frac{1}{\sqrt{2E}}\left(\begin{array}{c}1\\ 0\end{array}\right)
$ and likewise for dotted spinors.} 
Notice that spinors $\lambda_\alpha(p)$, $\rho_\alpha(p)$,  $\lambda_{\dot\alpha}(p)$, and $\rho_{\dot\alpha}(p)$ are now nicely defined up to $ISO(2)$ transformations 
 \begin{equation}
\label{eq:iso2equivalenceclass}
|i\rangle\to w |i\rangle\ ,\quad |\mathmybb{i}\rangle \to w^{-1}|\mathmybb{i}\rangle + 2i \alpha_{+}w^{-1}|i\rangle\ ,\quad [i|\to w^{-1} [i|\ ,\quad [\mathmybb{i}| \to w [\mathmybb{i}|  -2i\alpha_{-} w [i| \, ,
\end{equation}
which leave invariant \eqref{eq:momentum-spinor} and \eqref{eq:defrho}. The action of Lorentz $SL(2,\mathbb{C})\times SL(2,\mathmybb{C})$ on a particular choice $(\lambda_\alpha(p),\rho_\alpha(p))$ and $(\lambda_{\dot\alpha}(p),\rho_{\dot\alpha}(p))$ obeying \eqref{eq:momentum-spinor} and \eqref{eq:defrho}, returns $(\lambda_\alpha(\Lambda p),\rho_\alpha(\Lambda p))$ and $(\lambda_{\dot\alpha}(\Lambda p),\rho_{\dot\alpha}(\Lambda p))$ up to an ISO(2) transformation \eqref{eq:iso2equivalenceclass}. Therefore, the Lorentz-invariance constraint \eqref{eq:lorentz} is satisfied if amplitudes  are $SL(2,\mathbb{C})\times SL(2,\mathbb{C})$-invariant  functions of bi-spinors $(\lambda_\alpha^i,\rho_{\alpha}^i)$ and $(\lambda_{\dot\alpha}^i,\rho_{\dot\alpha}^i)$, which carry an ISO(2) representation for each CSP.

While translations are still realized by multiplication $P_{\alpha\dot\alpha}=\lambda_\alpha \lambda_{\dot\alpha}$ on this space, the Lorentz generators (for each particle) are \begin{equation}
J_{\alpha\beta}=i\lambda_{(\alpha}\frac{\partial}{\partial\lambda^{\beta)}}+i\rho_{(\alpha}\frac{\partial}{\partial\rho^{\beta)}}\ ,\quad J_{\dot{\alpha}\dot{\beta}}=i\lambda_{(\dot{\alpha}}\frac{\partial}{\partial\lambda^{\dot{\beta})}}+i\rho_{(\dot{\alpha}}\frac{\partial}{\partial\rho^{\dot{\beta})}}\ ,
\end{equation}
so that Pauli-Lubanski squared is the following functional:
\begin{equation}
\left(W_\mu W^\mu\right)_i=\frac{1}{2}\left(W_{\alpha\dot\alpha}W^{\alpha\dot\alpha}\right)_i =-\langle \lambda^i \rho^i \rangle [\lambda^i \rho^i] \lambda^i_{\alpha}\frac{\partial}{\partial\rho^i_\alpha} \lambda^i_{\dot\alpha}\frac{\partial}{\partial\rho^i_{\dot\alpha}}=
-\langle i \mathmybb{i} \rangle [i \mathmybb{i}]\, |i\rangle\frac{\partial}{\partial|\mathmybb{i}\rangle} |i]\frac{\partial}{\partial|\mathmybb{i}]}\ ,
\end{equation}
acting on the space of functions of bi-spinors. 
We recall that $W_\mu^2=-\mathbb{W}_{+} \mathbb{W}_{-}$. Thus, the differential operators $\mathbb{W}_{\pm}$, $\mathbb{H}$ acting on this space are easily extracted
\begin{equation}
\label{eq:iso2genbispinorsforapp}
    \mathbb{W}^{-}_i=\langle i \mathmybb{i} \rangle |i\rangle \frac{\partial}{\partial|\mathmybb{i}\rangle}\ ,\qquad \mathbb{W}^{+}_i= [i \mathmybb{i}]  |i] \frac{\partial}{\partial|\mathmybb{i}]}\ , \qquad
    \mathbb{H}_i= -\frac{1}{2}\left( |i\rangle \frac{\partial}{\partial |i\rangle} -|\mathmybb{i}\rangle \frac{\partial}{\partial |\mathmybb{i}\rangle} - |i] \frac{\partial}{\partial |i]}+|\mathmybb{i}] \frac{\partial}{\partial |\mathmybb{i}]}\right)\ .
\end{equation}
Indeed, they satisfy the $ISO(2)$ algebra \eqref{eq:iso2algebra}. The $\mathbb{W}^{-}$ annihilates $\lambda_\alpha$ and lowers $\rho_\alpha \to \propto \lambda_{\alpha}$, whereas $W^+$ annihilates $\lambda^{\dot\alpha}$ and  raises $\rho^{\dot\alpha} \to \propto \lambda^{\dot\alpha}$. Finally the $\mathmybb{W}^{\mp}$ have helicity $\mp$ respectively.
Labelling explicitly the amplitudes by  $\theta_i$-angles associated with the states, the amplitudes must solve  
\begin{equation}
\label{eq:iso2differentialOp}
\mathbb{W^{\pm}}_i\mathcal{M}_{\theta_1\ldots \theta_n}=\mu_i^{\pm}\mathcal{M}_{\theta_1\ldots \theta_n}\ ,\qquad 
\mathbb{H}_i \mathcal{M}_{\theta_1\ldots \theta_n}=i\frac{\partial}{\partial\theta_i}\mathcal{M}_{\theta_1\ldots \theta_n}\ ,
%
\qquad  \forall i=1,\dots,n\ . 
\end{equation}

Finally, since amplitudes are functions of invariant contractions only, these equations are equivalent to the following set of differential equations\footnote{\label{footnote:SchurLemma} Since $[i \mathmybb{i}]$ and $\langle i \mathmybb{i}\rangle $ are both Lorentz and little-group invariant, it may be sometimes practical to fix their normalization to \textit{e.g.} $-1$ and $+1$ respectively. In fact, $\rho$-dilation operators $D= \rho_\alpha \frac{\partial}{\partial_{\rho_\alpha}}$ and $\widetilde{D}= \rho_{\dot\alpha} \frac{\partial}{\partial_{\rho_{\dot\alpha}}} $ commute with all elements of little-group and Poincar\'e algebra so that by Schur's lemma they act as multiple of the identity on irreps of this larger group. Writing each amplitude as sum of irreps of $D$ and $\widetilde{D}$,  rescaling $\rho_{\alpha}$ and $\rho_{\dot{\alpha}}$ corresponds to multiply each of those terms by a rescaling factor (see appendix~\ref{appendix:rhospinors})  which we can interpret effectively as changing coupling constants. This is analogous to the unphysical phases carried by the variables $m$ and $\widetilde{m}$ for the massive spinors of reference~\cite{Arkani-Hamed:2017jhn}.} $\forall i=1,\ldots n$: 
\begin{subequations}
\label{eq:iso2differential1p}
\begin{align}
\label{eq:iso2differential1pWmin}
& \langle i \mathmybb{i} \rangle \sum_{j=1}^n\left(  \langle j i\rangle \frac{\partial}{\partial\langle j \mathmybb{i}\rangle} +  \langle \mathmybb{j} i\rangle \frac{\partial}{\partial\langle \mathmybb{j} \mathmybb{i}\rangle} + \ldots \right)\log\mathcal{M}_{\theta_1\ldots\theta_n} = \mu^{-}_i \ ,\\
\label{eq:iso2differential1pWplus}
& [i \mathmybb{i}] \sum_{j=1}^n\left(  [ j i] \frac{\partial}{\partial [ j \mathmybb{i}]} +  [ \mathmybb{j} i] \frac{\partial}{\partial [ \mathmybb{j} \mathmybb{i}]}+\ldots \right)\log\mathcal{M}_{\theta_1\ldots\theta_n} =  \mu^{+}_i \ ,\\
\label{eq:iso2differential1pH}
& \mathrm{Exp}\left(-i\omega\mathbb{H}_i\right)\mathcal{M}_{\theta_1\ldots \theta_n}=\mathcal{M}_{\theta_1\ldots(\theta_i+\omega)\ldots \theta_n}\ ,
\end{align}
\end{subequations}
 where we explicitly display differentiation w.r.t. massless spinors while ellipsis contain as well derivatives w.r.t. massive spinors, such as $\langle \boldsymbol{j}^I i\rangle \frac{\partial}{\partial\langle \boldsymbol{j}^I \mathmybb{i}\rangle}$ for $\mathbb{W}_{-}$, whenever present. 
The amplitudes in these differential equations are functions of  massless and massive spinors that may appear in the problem
\begin{equation}
\mathcal{M}_{\theta_1\ldots \theta_n}=\mathcal{M}_{\theta_1\ldots \theta_n}\left( \langle i j\rangle, \langle i \mathmybb{j}\rangle, \langle \mathmybb{i} \mathmybb{j}\rangle, \langle \boldsymbol{i} j\rangle, \langle \boldsymbol{i} \mathmybb{j}\rangle, \langle  \boldsymbol{i}{j}\rangle, [i j], [ \mathmybb{i} j],\ldots  \right)\ , 
\end{equation}
but typically momentum conservation, on-shell conditions, and Schouten identities reduce the number of independent contractions to be considered. 

We observe that the space of functions of spinors can actually be larger than the one we considered so far. It can contain as well complex conjugate spinors that carry complex conjugate irreps, within the complexified Lorentz group $SL(2,\mathbb{C})\times SL(2,\mathbb{C})$.  We study amplitudes resulting from this extended space of functions in Section~\ref{sec:non-analytic}. Until then, we restrict to functions that are analytic in the spinors.

\subsection{Three-point Amplitudes}
\label{sec:All3pts}

In this section, we present first some simple fully worked-out solutions to the kinematic constraints \eqref{eq:iso2differential1p} for three-point amplitudes involving CSPs, in preparation for the general $n$-point amplitude presented in the next sections. 

\subsubsection{2-CSP and 1 massive spin-$j$}
\label{subsec:2CSP1Massive}

This simple example covers several of the interesting general features that CSP amplitudes display. It is moreover interesting on its own to extract the partial-wave (PW) expansion \cite{Jacob:1959at} of a general 4-CSP amplitude, as it can be interpreted as the Poincar\'e Clebsh-Gordan coefficients (CGCs) for amplitudes with external CSPs, following \cite{Arkani-Hamed:2017jhn,Jiang:2020rwz,Shu:2021qlr}. This is discussed in Section~\ref{subsec:pw}.

We assign particles $i=1, 2$ to be CSPs, and $i=3$ the particle of mass $m=m_3$. Momentum conservation $|\mathbf{3}^I\rangle [\mathbf{3}_I |=-|1\rangle [1|-|2\rangle [2|$, on-shell conditions $\langle12\rangle[21]=m_3^2$, and the Schouten identities\footnote{We remind that $\langle i \mathmybb{i}\rangle$ and $[ i \mathmybb{i}]$ are ISO(2) invariant set to constants. Moreover, the Schouten identity $\langle \mathmybb{1} \mathmybb{2} \rangle \langle 1 2\rangle + \langle \mathmybb{1} 1 \rangle \langle \mathmybb{2} 2\rangle + \langle \mathmybb{1} 2 \rangle \langle \mathmybb{2} 1\rangle=0$, and analogous for square brackets, makes $\langle \mathmybb{1} \mathmybb{2} \rangle$ and $[\mathmybb{1} \mathmybb{2} ]$ linearly dependent w.r.t. the other contractions.} tell that the only non-trivial and independent contractions transforming under $ISO(2)$ are $\langle \mathmybb{1} 2 \rangle $, $\langle \mathmybb{2} 1 \rangle $, $[\mathmybb{1} 2 ] $, $[\mathmybb{2} 1 ] $, and either one between $\langle i j\rangle $ and $[i j]$.  
Therefore, the first two differential equations $W^{\pm}_i\mathcal{M}=~\mu^{\pm}_i$ for $i=1,2$ in \eqref{eq:iso2differential1pWmin} and \eqref{eq:iso2differential1pWplus} greatly simplify to 
\begin{subequations}
\begin{align}
\label{exampleFirstCSP}
 \langle 2 1\rangle \frac{\partial}{\partial\langle 2 \mathmybb{1}\rangle}\log\mathcal{M}_{\theta_1 \theta_2}=\frac{\mu^-_{1}}{\langle 1 \mathmybb{1}\rangle }\ , \qquad 
 [2 1 ] \frac{\partial}{\partial [ 2 \mathmybb{1}]}\log\mathcal{M}_{\theta_1 \theta_2}=\frac{\mu^+_{1}}{[ 1 \mathmybb{1}] }\ ,  \\ 
\langle 1 2\rangle \frac{\partial}{\partial\langle 1 \mathmybb{2}\rangle}\log\mathcal{M}_{\theta_1 \theta_2}=\frac{\mu^-_{2}}{\langle 2 \mathmybb{2}\rangle}\ ,
 \qquad 
[ 1 2] \frac{\partial}{\partial [ 1 \mathmybb{2}]}\log\mathcal{M}_{\theta_1 \theta_2}=\frac{\mu^+_{2}}{[ 2 \mathmybb{2}]} \ .
\end{align}
\end{subequations}
Their solutions are
\begin{equation}
\label{eq:simplest3pt}
\begin{split}
    \M_{\theta_1 \theta_2} 
    =\exp\left(\frac{\mu_1^{-}}{\langle 1 \mathmybb{1} \rangle} \frac{\langle 2 \mathmybb{1}\rangle }{\langle 2 1\rangle} + \frac{\mu_1^{+}}{[1 \mathmybb{1} ]} \frac{[ 2 \mathmybb{1}] }{[ 2 1]}\right) \exp\left(\frac{\mu_2^{-}}{\langle 2\mathmybb{2} \rangle} \frac{\langle 1 \mathmybb{2}\rangle }{\langle 1 2\rangle} + \frac{\mu_2^{+}}{[ 2 \mathmybb{2} ]} \frac{[ 1 \mathmybb{2}] }{[12]}\right)\widetilde{\mathcal{M}}_{\theta_1 \theta_2}\ ,
\end{split}
\end{equation}
where $\widetilde{\mathcal{M}}_{\theta_1 \theta_2}$ is a function of ordinary massless and massive spinors,  dependent on $\theta_{i=1,2}$ (and possibly on the $\mu_i$ and/or other constants). 

The final constraint from the helicity equation \eqref{eq:iso2differential1pH} is trivially solved by the exponential prefactors in \eqref{eq:simplest3pt}, \textit{e.g.} $\mu e^{-i\theta_1}\langle 2 \mathmybb{1}\rangle/\langle 2 1\rangle\to \mu e^{-i(\theta_1+\omega)}\langle 2 \mathmybb{1}\rangle/ \langle 2 1\rangle$ under $\mathrm{Exp(-i\omega \mathbb{H}_1)}$, because $|\mathmybb{i}\rangle$ and $|i\rangle$ have helicities $+1/2$ and $-1/2$, respectively.  The remaining non-trivial constraint imposed by \eqref{eq:iso2differential1pH} on $\widetilde{\mathcal{M}}_{\theta_1 \theta_2}$ can be solved in a similar vein, namely 
\begin{equation}
\label{eq:simplest3ptHE}
\widetilde{\mathcal{M}}_{\theta_1 \theta_2}=\sum_{\{h_i\} }\mathcal{M}_{h_1 h_2} e^{-ih_1 \theta_1 -ih_2 \theta_2}\ , 
\end{equation}
where $\mathcal{M}_{h_1 h_2}=\mathcal{M}_{h_1 h_2}(| i\rangle,  |j], |\mathbb{3}\rangle, |\mathbb{3}])$ are ordinary three-point amplitudes between two (ordinary) massless particles of helicity $h_{i=1,2}$ and a massive particle of spin-$j$. These are completely classified, see \textit{e.g.} \cite{Arkani-Hamed:2017jhn}, 
and therefore the problem is fully solved, see \eqref{eq:genera3pt2CSP1MassJ}. 

Finally, we highlight two features that are discussed in full generality in the following: {\it i)} the massless limit $m\to 0$ where all particles are degenerate is singular because $\langle 12\rangle[21]=m^2$; {\it ii)} the high-energy limit\footnote{For three-point amplitudes, this limit requires all the mass scales to be large (or equivalently taking $\mu \to 0$), see discussion of Section \ref{sec:highenergyAndStatistics}.} of CSPs just returns $\mathcal{M}_{\theta_1 \theta_2}\to \widetilde{\mathcal{M}}_{\theta_1 \theta_2}$, which Fourier-transformed back to helicity-basis gives the ordinary amplitudes $\mathcal{M}_{h_1 h_2}$ in \eqref{eq:simplest3ptHE}. Therefore, one can think of the three-point amplitude \eqref{eq:simplest3pt} as the IR deformation of a UV theory that had only ordinary massless particles of helicity $h_i$ in the three-point amplitude, with infinitely many helicities recoupling-in as the momenta are lowered down to $\mu$. For instance, choosing 
\begin{equation}
\label{eq:exampleMostlyPhotontoAxion}
\mathcal{M}_{h_1 h_2}= \delta_{h_1}^{-1} \delta_{h_2}^{-1} \langle 12 \rangle^2/f_{-} + \delta_{h_1}^{1} \delta_{h_2}^{1} [ 12 ]^2/f_{+}
\end{equation}
corresponds to two (neutral) CSPs coupled to a massive scalar $j=0$,  in a way that at high energy only helicities $\pm1$ remain coupled, with strength  set by decay  constants $f_{\pm}$. (Pairs of opposite helicities $\pm h$ remain at high energy because the action of CPT is closed on a single ---neutral--- CSP).
This high-energy limit is the same as produced by (a linear combination of) field theory interactions $F_{\mu\nu}F^{\mu\nu}\phi$ and $F_{\mu\nu}\widetilde{F}^{\mu\nu}\phi$ for Higgs-like and/or axion-like particles.   

Another example is the coupling of two (neutral) CSPs that in the high-energy limit reduces to the ``minimal coupling'' of $\pm 1$-helicity photons coupled to a massive $j=2$ particle, corresponding to the choice 
\begin{equation}
\mathcal{M}_{h_1 h_2}= \delta_{h_1}^{-1} \delta_{h_2}^{+1} \frac{\langle 1 \boldsymbol{3} \rangle^4}{\langle 12\rangle^2}/f_{-} + 
\delta_{h_1}^{+1} \delta_{h_2}^{-1} \frac{[ 1 \boldsymbol{3} ]^4}{[ 12]^2}/f_{+} \ , 
\end{equation}
where the interaction strength is set by decay constants $f_{\pm}$.  In a parity preserving theory $f_{-}=f_{+}$ and this is what it is obtained by coupling a massive spin-2 particle to the photon energy-momentum tensor.

The resulting general amplitude, which is useful for example for the PW-decomposition, is reported here for later convenience: 
\begin{subequations}
\label{eq:genera3pt2CSP1MassJ}
    \begin{align}
    \mathcal{M}_{\theta_1\theta_2} =&     \exp\left(\frac{\mu_1^{-}}{\langle 1 \mathmybb{1} \rangle} \frac{\langle 2 \mathmybb{1}\rangle }{\langle 2 1\rangle} + \frac{\mu_1^{+}}{[1 \mathmybb{1} ]} \frac{[ 2 \mathmybb{1}] }{[ 2 1]}\right) \exp\left(\frac{\mu_2^{-}}{\langle 2\mathmybb{2} \rangle} \frac{\langle 1 \mathmybb{2}\rangle }{\langle 1 2\rangle} + \frac{\mu_2^{+}}{[ 2 \mathmybb{2} ]} \frac{[ 1 \mathmybb{2}] }{[12]}\right)\widetilde{M}_{\theta_1 \theta_2}\ , \\
    \widetilde{M}_{\theta_1 \theta_2} =&  \sum_{h_1  h_2} c_{j h_1 h_2} \frac{e^{-i\theta_1 h_1-ih_2\theta_2}}{m^{3j+h_1+h_2-1}} 
    \langle 1 \boldsymbol{j}\rangle^{j+h_2-h_1} \langle 2 \boldsymbol{j}\rangle^{j+h_1-h_2}
    [12]^{j+h_1+h_2}\ .
\end{align}
\end{subequations}
with $\abs{h_1-h_2}\leq j$.
By using on-shell conditions these expressions can be reshuffled into different choices of angle and square brackets \cite{Arkani-Hamed:2017jhn}. 

\subsubsection{1-CSP, 1 massless particle and 1 massive spin-$j$}
\label{subsec:1CSP+1massless+1massive}

This example is a simple variation of the case study in previous Section~\ref{subsec:2CSP1Massive}  where now one of the two $\mu_i$ is set to zero, for instance $\mu_{2}=0$. For this reason, we highlight only its main steps. 

Particle 1 is the CSP, particle 2 is the ordinary massless particle of helicity $h_2$,  while particle 3 has mass $m=m_3$.
We can choose the non-trivial contractions with the black-board spinors to be $\langle\mathmybb{1} 2\rangle$ and $[\mathmybb{1}2]$.  The differential equations \eqref{eq:iso2differential1pWmin} and \eqref{eq:iso2differential1pWplus} for $\mathbb{W}^{\pm}$ reduce again to just \eqref{exampleFirstCSP}. Mutatis mutandis the solution is 
\begin{equation}
\label{Example3pt:1CSP-1massless-1massive}
\mathcal{M}_{\theta_1}=\exp\left(\frac{\mu_1^{-}}{\langle 1 \mathmybb{1} \rangle} \frac{\langle 2 \mathmybb{1}\rangle }{\langle 2 1\rangle} + \frac{\mu_1^{+}}{[1 \mathmybb{1} ]} \frac{[ 2 \mathmybb{1}] }{[ 2 1]}\right) \left(\sum_{h_i}\mathcal{M}_{h_1 h_2} e^{-ih_1 \theta_1}\right)
\end{equation}
where again $\mathcal{M}_{h_1 h_2}=\mathcal{M}_{h_1 h_2}(| i\rangle,  |j], |\mathbb{3}\rangle, |\mathbb{3}])$ are ordinary three-point amplitudes between two (ordinary) massless particles of helicity $h_{i=1,2}$ and a massive particle of spin-$j$. These are classified and the problem is fully solved. Notice again that the high-energy limit is trivially set by $\mathcal{M}_{h_1 h_2}$, and the massless limit $m_3\to0$ is singular because momentum conservation and on-shell conditions demand $\langle12\rangle[21]=m^2_3$. 

We note that adding black-board spinors $|\mathmybb{2}\rangle$ and $|\mathmybb{2}]$ for ordinary massless particle 2 is just equivalent to repeat the example in Section~\ref{subsec:2CSP1Massive} and take $\mu_2=0$. The solution collapses to \eqref{Example3pt:1CSP-1massless-1massive} after Fourier transforming, effectively setting $\theta_2=0$. 
This is actually a general fact: adding black-board spinors to ordinary massless particles $j$ can not enlarge (non-trivially) the space of solutions. The reason is that from  $\mathbb{W}^{\pm}_{j}\mathcal{M}=0$ it follows that the amplitude can at most carry the dependence on black-board spinors only via the choice of normalizations $\langle j \mathmybb{j}\rangle$ and $| j \mathmybb{j}]$.

\subsubsection{1-CSP and 2 massive spin $j_1$, $j_2$}
\label{subsec:1CSP+2massive}

We choose for this example particle 3 as the CSP, while particles 1 and 2 carry spin  $j_1$, $j_2$ and masses $m_1$, $m_2$. 
From playing with previous examples, it is not hard to guess the solution to the $ISO(2)$ functional problem \eqref{eq:iso2differential1p}, namely 
\begin{equation}
\label{eq:1CSP2massive}
\begin{split}
    \M_{\theta_3} 
    =\exp\left(\frac{\mu^{+}_3}{\sqr{3}{\mathmybb{3}}} \frac{\langle 3 | p_1 
    | \mathmybb{3} ]}{\langle 3| p_1 |3]    } + \frac{\mu^{-}_3}{\agl{3}{\mathmybb{3}}} \frac{\langle \mathmybb{3} | p_1 
    | 3 ]}{ \langle 3| p_1 |3]   }\right) \left(\sum_{h_3} \mathcal{M}_{h_3} e^{-ih_3\theta_3}\right)\ ,
\end{split}
\end{equation}
where
$\mathcal{M}_{h_3}$ is again an ordinary amplitude between one massless particle of helicity $\{ h_3\}$ and the two massive spinning particles. Its choice fixes the high-energy limit of the CSP to be the same as the one of a massless particle of helicity $h_3$ coupled to those spins. Finally, the limit where two particles are degenerate in mass is again singular because momentum conservation and on-shell condition enforce $m_2^2-m_1^2=\langle 3|p_1 |3]\to 0$ and the amplitude has an essential singularity. We checked that there are no other solutions to the equations~\eqref{eq:iso2differential1pWmin}~and~\eqref{eq:iso2differential1pWplus} in the case $m_1=m_2$, since the only independent contraction is $\langle \mathmybb{3} | p_1| \mathmybb{3} ]$.

This amplitude may be used to describe the excitation of a composite particle or extended system that has more energy levels, through the absorption or the emission of CSPs.

\subsubsection{3-CSP and the mass-splitting selection rule}
\label{sec:masssplittingselection}

 A striking observation of this subsection is that on-shell three-massless interactions with at least one CSP are always kinematically forbidden.\footnote{This conclusion can be avoided by extending the space of amplitudes by including non-analytic functions of spinors and their complex conjugate.  We defer the discussion of exotic non-analytic three-point amplitudes to Section~\ref{sec:non-analytic}.}

We remind the reader that three-point massless kinematics requires the vanishing of all Mandelstam invariants 
\begin{equation}
\label{eq:real_3pt_kin}
    s_{i j} = \agl{i}{j} \sqr{j}{i} = 0 \quad \forall (i,j)\ .
\end{equation}
Complex momenta allow for a solution of this constraint where either 
\begin{equation}
    \sqr{i}{j} = 0 \quad \forall (i,j)\ , \quad \text{or} \quad \agl{i}{j} = 0 \quad \forall (i,j)\ ,
\end{equation}
to which we usually refer as \textit{holomorphic} and \textit{anti-holomorphic} configurations, respectively.
While complex kinematics is enough for ordinary massless particles to admit non-trivial three-point amplitudes, this is not the case for CSPs.  
In order to show this, we consider a holomorphic configuration, such that the only non-vanishing square brackets are $\sqr{i}{\mathmybb{j}}$. 
Therefore, the $\mathbb{W}^+$-differential equations \eqref{eq:iso2differential1pWplus} demand
\begin{equation}
\sum_{j\neq i}^3   [ \mathmybb{j} i] \frac{\partial}{\partial [ \mathmybb{j} \mathmybb{i}]} \mathcal{M}_{\theta_1\ldots\theta_n}  =\frac{\mu^+_i}{[i\mathmybb{i}]}\mathcal{M}_{\theta_1\theta_2\theta_3}\ , \quad \forall i=1,2,3\ .
\end{equation}
Enforcing momentum conservation  and using $\langle k i \rangle [i \mathmybb{j}]+\langle k j\rangle [j \mathmybb{j}]=0$ for $i\neq j\neq k$, we can express $[i \mathmybb{j}]$ in terms of angle bracket and constant $[j \mathmybb{j}]$. The resulting system has no solutions unless all $\mu_i=0$. An equivalent argument applies to the anti-holomorphic configuration. 

The same conclusion can be reached by guessing the solution (see also \eqref{eq:CS_kinematics})
\begin{equation}
\prod_{i=1}^3 \exp\left(\frac{\mu_i^{-}}{\langle i \mathmybb{i} \rangle} \frac{\langle \xi \mathmybb{i}\rangle }{\langle \xi i\rangle} + \frac{\mu_i^{+}}{[i \mathmybb{i} ]} \frac{[ \xi \mathmybb{i}] }{[ \xi i]}\right)\widetilde{\mathcal{M}} \ ,
\end{equation}
for some pair of spinors $|\xi\rangle$ and $|\xi]$. As soon as one requires these auxiliary spinors to be a non-vanishing linear combination of the spinors of the problem,\footnote{We will discuss possible way to relax this assumption in Section~\ref{sec:backgrounds}.} and demand momentum conservation, either one between $[\xi i]$ or $\langle \xi i\rangle$ has to vanish, and so the solution diverges unless all $\mu_i=0$.

Taken at face value, this conclusion, along with the results of previous subsections, implies that no on-shell three-point amplitude exists (within the assumptions made) among at least one CSP and two other particles of degenerate mass.\footnote{This conclusion cannot be avoided by the so-called $x$-factor in the classification of \cite{Arkani-Hamed:2017jhn}, as it is already
appearing in the numerator of the exponents of the kinematic phases as $\langle \xi \mathmybb{i}\rangle/\langle i \mathmybb{i}\rangle$, but LG covariance forces the denominator to be $\langle \xi i\rangle$, which is always vanishing.} We refer to this condition as {\it the  ``mass-splitting selection rule''}. \\ 
This finding is consistent with previous literature that observed singular soft-factors in the degenerate mass limit~\cite{Schuster:2013pxj,Schuster:2013vpr}.  Moreover, a specific model of CSP coupled to two scalars of unequal masses  has been explicitly studied in reference~\cite{Bekaert:2017xin}.  The mass-splitting selection rule we have found  represents an universal selection rule, not one specific to a given model or to a kinematical limit.

The mass-splitting selection rule has interesting consequences: 
\begin{enumerate}
    \item No \textit{on-shell} coupling of CSP to (massless) gravitons is possible. 
    Similarly, CSPs cannot be coupled on-shell to photons. 
    \item CSPs have no on-shell three-point self-interactions, \textit{e.g.} no CSP-like gravitons nor non-abelian (massless) gauge theory can be recovered, on-shell, in the high-energy limit. 
    \item Mass-preserving on-shell coupling to matter fields is forbidden and the CSP can't reduce to an on-shell graviton in the UV limit. 
\end{enumerate}

The only on-shell three-point configurations that are consistent with all assumptions are those that we discussed in previous subsections.
We also note that the mass-splitting selection rule uniquely selects, despite the use of complex kinematics,  on-shell 3-point functions that survive in the limit of physical kinematics: the decay of the heaviest particle into at least one CSP and a lighter particle. 

However, we emphasise that in certain higher-point amplitudes, that satisfy factorisation into consistent three-point amplitude (hence respecting the mass-splitting selection rule), the limit of degenerate masses does actually exist. This arises typically when one of the would-be degenerate particles appears off-shell in some higher n-point amplitude, see \textit{e.g.} Section~\ref{Sec:Photon-Gravity} where we couple CSPs to exchanged massive gravitons and take the limit of massless gravitons at the end.  

Moreover,  even though some three-point amplitudes may vanish on-shell because of the mass-splitting selection rule, some of them admit a mass deformation which is suggestive that off-shell quantities, such as expectation values of certain operators, actually exist. 
For example, we can consider form factors for bosonic CSPs, such as conserved currents
\begin{equation}
\label{eq:Jmumatrixelement}
\langle 0| J^{\mu}(0)| 1^{\theta_1} 2^{\theta_2} \rangle= \left[ G(q^2) (p_1 -p_2)^\mu+\ldots \right] \exp\left(\frac{\mu_1^{-}}{\langle 1 \mathmybb{1} \rangle} \frac{\langle 2 \mathmybb{1}\rangle }{\langle 2 1\rangle} + \frac{\mu_1^{+}}{[1 \mathmybb{1} ]} \frac{[ 2 \mathmybb{1}] }{[ 2 1]}\right) \exp\left(\frac{\mu_2^{-}}{\langle 2\mathmybb{2} \rangle} \frac{\langle 1 \mathmybb{2}\rangle }{\langle 1 2\rangle} + \frac{\mu_2^{+}}{[ 2 \mathmybb{2} ]} \frac{[ 1 \mathmybb{2}] }{[12]}\right)\ ,
\end{equation}
and the energy-momentum tensor
\begin{align}
\nonumber
\langle 0| T^{\mu\nu}(0)| 1^{\theta_1} 2^{\theta_2} \rangle=&  \left[F_0(q^2)(p_1- p_2)^\mu (p_1- p_2)^\nu +F^\prime_0(q^2)P^{\mu\nu}(q) + F_{1}(q^2)\langle 1| \sigma^\mu |2] \langle 1| \sigma^\nu| 2]  e^{i(\theta_1-\theta_2)}+\ldots \right]\\
\label{eq:Tmunumatrixelement}
 &\times \exp\left(\frac{\mu_1^{-}}{\langle 1 \mathmybb{1} \rangle} \frac{\langle 2 \mathmybb{1}\rangle }{\langle 2 1\rangle} + \frac{\mu_1^{+}}{[1 \mathmybb{1} ]} \frac{[ 2 \mathmybb{1}] }{[ 2 1]}\right) \exp\left(\frac{\mu_2^{-}}{\langle 2\mathmybb{2} \rangle} \frac{\langle 1 \mathmybb{2}\rangle }{\langle 1 2\rangle} + \frac{\mu_2^{+}}{[ 2 \mathmybb{2} ]} \frac{[ 1 \mathmybb{2}] }{[12]}\right) \,, 
\end{align}
where $q=p_1+p_2$ and $P^{\mu\nu}=q^\mu q^\nu-\eta^{\mu\nu}q^2$. In Section~\ref{Sec:Photon-Gravity} we provide an example where the coupling to massive spin-2 particle can be thought of as the insertion of these energy-momentum tensor matrix elements.  In particular, the  choice made there, $F_{h}=\mathrm{const}\times\delta_{h}^{\pm 1}$ while all other form factors vanish, corresponds to the case of a gravitationally minimally-coupled photon, in the $\mu\to 0$ limit. 

\subsection{$n$-point Amplitudes}
\label{subsec:npointAmpl}

We have seen in the previous sections that amplitudes are solutions of a linear differential problem defined by the little-group of each particle as realized in the space of spinors, \eqref{eq:iso2differential1p}.  
Solving these constraints for $n$-point amplitudes is simplified by the existence of the following family of solutions 
\begin{equation}
    X_j^+(p^{+}_{j})= e^{\frac{\mu^+_{j}}{\sqr{j}{\mathmybb{j}}} \frac{\langle j | p^+_{j} | \mathmybb{j} ]}{\langle j | p^+_{j} | j ]}}\ ,\qquad 
    X_j^-(p^{-}_{j})= e^{\frac{\mu_{j}^{-}}{\agl{j}{\mathmybb{j}}} \frac{\langle \mathmybb{j} | p^-_{j} | j ]}{\langle j | p^-_{j} | j ]}}\ ,\qquad 
    \mathbb{W}^{\pm}_j X^{\pm}_j= \mu^ {\pm}_{j} X^{\pm}_j\ ,\qquad 
    \mathbb{H}_j X^{\pm}_{j}=i\frac{\partial}{\partial\theta_j} X_{j}^{\pm}\ , 
%
\end{equation}
for any (non-singular) choice of $p^{+ \mu}_{j}$ (space-, light- or time-like vector).
Thus, if such a choice exists, we can write
\begin{equation}
\label{eq:CS_kinematics}
    \mathcal{M}_n =\sum_{{p_j^{\pm}}} \exp\left(\sum\limits_{j=1}^C \frac{\mu_{j}^{+}}{\sqr{j}{\mathmybb{j}}} \frac{\langle j | p^+_{j} | \mathmybb{j} ]}{\langle j | p^+_{j} | j ]} + \frac{\mu_{j}^{-}}{\agl{j}{\mathmybb{j}}} \frac{\langle \mathmybb{j} | p^-_{j} | j ]}{\langle j | p^-_{j} | j ]}\right) \widetilde{\mathcal{M}}^n_{\theta_1\ldots\theta_C} \ ,
\end{equation}
with $C$ the number of continuous-spin particles, and $p^{\pm}_{i}$ are analytic functions of bi-spinors such that on real kinematic $p^{-}_i\to p^{+\,*}_i$, 
and
\begin{equation}
\label{eq:helicity_constraint_Mtilde}
    \mathbb{W}^{\pm}_j \widetilde{\mathcal{M}}^n_{\theta_1\ldots\theta_C} = 0 \ , \qquad 
    \mathbb{H}_{j} \widetilde{\mathcal{M}}^n_{\theta_1\ldots \theta_C} = i \frac{\partial}{\partial \theta_j} \widetilde{\mathcal{M}}^n_{\theta_1\ldots \theta_C} \ .
\end{equation} 
Occasionally, we do not display all labels and indices of this expression. 

First, we enforce the $\mathbb{H}_j$-constraint in \eqref{eq:helicity_constraint_Mtilde}.  It is promptly solved by a sum of harmonics weighed  by amplitudes of assigned helicities $h_i$ (and possibly other particles whose quantum numbers are not displayed): 
\begin{equation}
\label{eq:generalNpointWtilde}
\widetilde{\mathcal{M}}_n = \sum_{\{h_i\}} e^{-i \sum_i \theta_i h_i} \mathcal{M}^n_{h_1 \dots  h_C} \qquad \text{with} \quad \mathbb{H}_j \mathcal{M}^n_{h_1 \dots  h_C} = h_j\, \mathcal{M}^n_{h_1 \dots  h_C}\ .
\end{equation}
Therefore, the $\mathbb{W}^{\pm}_j$-constraints in \eqref{eq:helicity_constraint_Mtilde} tell us that  $\mathcal{M}^n_{h_1 \dots  h_C}$ solves literally the same constraints that apply for ordinary massless particles. For this reason, it can be shown ---see appendix \ref{appendix:rhospinors}--- that $\mathcal{M}^n_{h_1 \dots  h_C}$ depends on all spinors but the black-board spinors $\rho$'s, with the exception of the trivial contractions $\langle i \mathmybb{i}\rangle$ and $[i \mathmybb{i}]$, which are instead allowed. As discussed in  footnote~\ref{footnote:SchurLemma}, they are just constants set by the choice of normalization, and their dependence in $\mathcal{M}^n_{h_1 \dots  h_C}$  is left understood in the following.   

As explained in Section~\ref{sec:highenergyAndStatistics}, the high-energy limit of the $n$-point amplitude is fully determined by the ordinary amplitudes $\mathcal{M}^n_{h_1, \dots , h_C}$.

    It may be convenient to introduce in the following a more compact notation for the covariant exponents of the CSPs, introducing the polarisations
    \begin{equation}
    \label{footnote:epsilonDefinition}
        \epsilon_{j \alpha \dot{\alpha}}^+ = -i e^{+ i \theta_j} \frac{\lambda_{j \alpha} \rho_{j \dot{\alpha}}}{\sqr{j}{\mathmybb{j}}}\ , \quad \epsilon_{j \alpha \dot{\alpha}}^- = -i e^{- i \theta_j} \frac{\rho_{j \alpha} \lambda_{j \dot{\alpha}}}{\agl{j}{\mathmybb{j}}}\ , \quad \text{and} \quad \epsilon_{j \alpha \dot{\alpha}} = \epsilon_{j \alpha \dot{\alpha}}^+ + \epsilon_{j \alpha \dot{\alpha}}^-\ ,
    \end{equation}
    where we emphasise that $\epsilon_j$ is real in real kinematics. The amplitude is thus written as
    \begin{equation}
        \mathcal{M}_n = \sum_{{p_j^{\pm}}}\exp\left(i \sum\limits_{j=1}^C \mu_{j} \frac{\epsilon^+_j \cdot p^+_{j}}{k_j \cdot p^+_{j}} + \mu_{j}  \frac{\epsilon^-_j \cdot p^-_{j}}{k_j \cdot p^-_{j}} \right) \widetilde{\mathcal{M}}^n_{\theta_i} \overset{p_+^\mu \in \mathbb{R}^4}{=} \sum_{{p_j}}\exp\left(\sum\limits_{j=1}^C i\, \mu_{j} \frac{\epsilon_j \cdot p_{j}}{k_j \cdot p_{j}} \right) \widetilde{\mathcal{M}}^n_{\theta_i}\ .
    \end{equation}
    Notice that in real kinematics $\epsilon_j$ coincides with the $\eta(\phi)$ of reference~\cite{Schuster:2023jgc}, appearing in a similar exponential structure. Nevertheless, our definition naturally provides an analytic continuation to complex kinematics, while in reference~\cite{Schuster:2023jgc} $\eta(\phi)$ is kept real even when complexifying the Mandelstam variables through an $i\epsilon$ prescription. In this sense the $\eta(\phi)$ is more closely connected to the non-analytic amplitudes discussed in Section~\ref{sec:non-analytic}. 

\subsection{High-energy limit and spin-statistics for CSP}
\label{sec:highenergyAndStatistics}
We have mentioned already a few times that the high-energy limit of amplitudes with CSPs is determined by the ordinary $\mathcal{M}^n_{h_1\ldots h_C}$ amplitudes which are found inside the $\widetilde{\mathcal{M}}^n_{\theta_1\ldots \theta_C}$ that multiplies the exponential solutions of the $ISO(2)$ differential equations, see \textit{e.g.} \eqref{eq:simplest3pt},  \eqref{eq:simplest3ptHE}, \eqref{eq:CS_kinematics} and \eqref{eq:generalNpointWtilde}.

The form of the exponential factors suggests that the limit in which the energy\footnote{For three-point amplitudes (and, in some instances, for higher-point amplitudes with exchanges of CSPs), we need to require also the masses (or their differences) to be larger than $\mu_i$.} of the CSP is larger than $\mu_i$ is effectively equivalent to $\mu_i\to0$. By Fourier transforming back to the helicity basis, in this limit one recovers the amplitudes $\mathcal{M}^n_{h_1\ldots h_C}$ for ordinary massless particles of helicities determined by the support of the sum~\eqref{eq:generalNpointWtilde}.

For $E_i\gg \mu_i$, all helicities $h_i$ disappear from the amplitude except for those contained inside $\widetilde{\mathcal{M}}$. We will see this phenomenon in an explicit computation in Section~\ref{Sec:Photon-Gravity}.
Whenever $\widetilde{\mathcal{M}}$ contains only amplitudes that correspond to a sensible massless particle of helicity $h$, we refer to it as \textit{mostly-helicity} $h$ CSPs. For a CSP which  is its own anti-particle, \textit{i.e.} closed under the action of CPT, both pairs of helicities $\pm h$  must appear inside $\widetilde{\mathcal{M}}$. Conversely, a charged CSP, \textit{i.e.} mapped by CPT to a distinct anti-CSP, can be mostly one helicity $h$ only (like in a chiral theory), and its anti-particle would thus be mostly helicity $-h$. Notice that one could even consider mixed (or rather unified) cases, e.g. featuring a CSP with mostly-$(|h|=2)$ coupling to all particles and additionally a mostly-$(|h|=1)$ coupling to charged particles only.

In the opposite ---soft--- limit $E_i\lesssim \mu_i$ all helicities re-couple and the exponential prefactor is wildly oscillating, greatly changing the soft behaviour of a mostly-$h$ CSP w.r.t. its exact $\mu_i=0$ counterpart.

An interesting observation is that the exponential prefactors for the three-point function in \eqref{eq:simplest3pt} are even w.r.t. the exchange of the two identical CSPs in the same $\theta_i$ configuration. Since the solution of these three-point amplitudes is unique and even w.r.t. the exchange of the label of the two identical CSPs, the fermionic or bosonic statistics is determined by the UV behaviour of this amplitude. That is, CSPs in three-point amplitudes are bosons or fermions depending on whether the UV amplitude associated ---$\mathcal{M}_{h_1 h_2}$--- is even or odd w.r.t. the exchange of $1\leftrightarrow 2$, respectively.

\section{Examples}
\label{sec:ExamplesGeneral}

In this section, we discuss some paradigmatic examples of four-point amplitudes involving CSPs ---both on-shell and off-shell---  that respect several welcome properties:  {\it i)} Lorentz invariance and little-group covariance,  {\it ii)} factorisation in three-point amplitudes\footnote{In particular, we require each three-point amplitude to depend only on the spinors associated to the momenta of the corresponding vertex. Relaxing this condition would represent a significant departure from locality, which we find difficult to justify. Hence, we have chosen not to pursue this latter possibility further.},  {\it iii)} good high-energy behaviour at $E_i\gg \mu_i$ such that only one $\pm h_i$-helicity remains interacting, {\it iv)} crossing symmetry.  

 Examples that relax these properties are discussed in  Section~\ref{sec:weakening-Assumptions}.

\subsection{CSPs and Euler-Heisenberg}
\label{Sec:Euler-Heisenberg}

We consider the theory of a mostly-$(|h|=1)$ CSP coupled to a spin-$j=0$ particle. We later integrate the latter to obtain an Euler-Heisenberg-like theory of CSPs.

The fundamental building block is the three-point amplitude for 2 CSPs and one massive particle we discussed in Section~\ref{subsec:2CSP1Massive}. 
The only helicity structures allowed in the UV for the mostly-($|h|=1$) CSP is the ``helicity-flipping'' ones, namely $h_1=h_2=+1$ or $h_{1}=h_2=-1$, as reported in \eqref{eq:exampleMostlyPhotontoAxion}. Restricting for simplicity to a parity-preserving theory, $f_+=f_-\equiv f$,  we thus have: 
\begin{equation}
\begin{split}
    \M(1^{\theta_1}2^{\theta_2}3^{j=0})&=\frac{1}{f}\exp{\frac{\mu}{\langle 1\mathmybb{1}\rangle} e^{-i\theta_1}\frac{\langle \mathmybb{1}2\rangle}{\langle 12\rangle}+\frac{\mu}{[1\mathmybb{1}]} e^{i\theta_1}\frac{[\mathmybb{1}2]}{[12]}}\exp{\frac{\mu}{\langle 2\mathmybb{2}\rangle} e^{-i\theta_2}\frac{\langle \mathmybb{2}1\rangle}{\langle 21\rangle}+\frac{\mu}{[2\mathmybb{2}]} e^{i\theta_2}\frac{[\mathmybb{2}1]}{[21]}}\cdot\\
    &\cdot\left[\langle 12\rangle^2 e^{i(\theta_1+\theta_2)}+[12]^2 e^{-i(\theta_1+\theta_2)}\right]\ .
\end{split}
\end{equation}

We can then write an ansatz for the 4-point amplitude with external CSPs satisfying factorisation in the $s_{12}$-, $s_{13}$- and $s_{14}$-channel:
\begin{equation}
\begin{split}
\M(1^{\theta_1}2^{\theta_2}3^{\theta_3}4^{\theta_4})&=-\frac{1}{f^2(s_{12}-m^2)}e^{i\mu\frac{\epsilon_1\cdot p_2+\epsilon_2\cdot p_1}{p_1\cdot p_2}}e^{i\mu\frac{\epsilon_3\cdot p_4+\epsilon_4\cdot p_3}{p_3\cdot p_4}}\cdot\\
&\cdot\left[\langle 12\rangle^2 e^{i(\theta_1+\theta_2)}+[12]^2 e^{-i(\theta_1+\theta_2)}\right]\cdot\left[\langle 34\rangle^2 e^{i(\theta_3+\theta_4)}+[34]^2 e^{-i(\theta_3+\theta_4)}\right]+\\
&+(2\leftrightarrow3)+(2\leftrightarrow4)\ .
\end{split}
\end{equation}
\begin{figure}[h!!]
    \centering
    \includegraphics{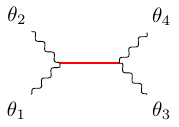}
    \caption{Euler-Heisenberg-like amplitude ($s$-channel).}
    \label{fig:eulerheisenberg}
\end{figure}
By construction the UV behaviour matches a standard theory of photons: when $s,t,u\gg\mu^2$ the exponentials are negligible and the trasformation to helicity basis allows to select separately each of the helicity structures.

Integrating out the massive particle we get the amplitude in the EFT for mostly-photon CSP-version of Euler-Heisenberg:
\begin{equation}
\label{eq:eulerheisenberg}
\begin{split}
\M_{\rm EFT}(1^{\theta_1}2^{\theta_2}3^{\theta_3}4^{\theta_4})&=\frac{1}{f^2m^2}e^{i\mu\frac{\epsilon_1\cdot p_2+\epsilon_2\cdot p_1}{p_1\cdot p_2}}e^{i\mu\frac{\epsilon_3\cdot p_4+\epsilon_4\cdot p_3}{p_3\cdot p_4}}\cdot\\
&\cdot\left[\langle 12\rangle^2 e^{i(\theta_1+\theta_2)}+[12]^2 e^{-i(\theta_1+\theta_2)}\right]\cdot\left[\langle 34\rangle^2 e^{i(\theta_3+\theta_4)}+[34]^2 e^{-i(\theta_3+\theta_4)}\right]+\\
&+(2\leftrightarrow3)+(2\leftrightarrow4)\ ,
\end{split}
\end{equation}
which is valid for $s_{ij}\ll m^2$, and where to make expression nicer we have used the polarizations $\epsilon_i$ as explained in equation~\eqref{footnote:epsilonDefinition}. 

\subsection{CSP exchange at tree-level}
\label{sec:CSPexchange}

In this subsection, we are interested in exploring the behaviour of amplitudes where interactions are mediated by an intermediate CSP. 

Since three-point interactions are subject to the mass-splitting selection rule discussed in Section~\ref{sec:masssplittingselection}, we consider a mass-changing interaction through the exchange of a mostly-scalar CSP at tree level. 
We may think of it as a two-level system with masses $m_i$ and $m^\prime_i$, $i=1,2$, that are coupled via a CSP with  three-point amplitude:
\begin{equation}
\label{eq:3pt_scalarsDeltam}
    \mathcal{M}_3 (1\,1^\prime\, k^{\theta})=\Lambda\exp\left(\frac{\mu^{+}}{\sqr{k}{\mathmybb{k}}} \frac{\langle k | p_1 | \mathmybb{k} ]}{\Delta m_1^2} + \frac{\mu^{-}}{\agl{k}{\mathmybb{k}}} \frac{\langle \mathmybb{k} | p_1 | k ]}{\Delta m_1^2}\right) \ ,
\end{equation}
where we have recast the three-point \eqref{eq:1CSP2massive} using the explicit mass-shell condition in terms of the mass splitting $\Delta m_i^2=m_i^{\prime\,2}-m_i^2$.  We consider the process $12\to1'2'$.
The four-point amplitude is computed by making an ansatz and probing it on its factorisation channels. In particular, this process admits only an $t$-channel exchange, and therefore we have\footnote{In this Section~\ref{sec:CSPexchange} we use the following definition of Mandelstam invariants: $s = (p_1+p_2)^2$, $t = (p_1+p_1^\prime)^2$ and $u = (p_1+p_2^\prime)^2$ .
}
\begin{equation}
\begin{split}
    \underset{t=0}{\text{Res}}\, \mathcal{M} (1\,1'\,2\,2') &= - \int_0^{2\pi} \frac{\dd\theta}{2\pi} \mathcal{M}_3 (1\,1'\,k^{\theta}) \mathcal{M}_3 (2\,2'\,(-k)^{\theta+\pi})\\
    & = - \Lambda^2 \int_0^{2\pi} \frac{\dd\theta}{2\pi} \exp\left(\frac{\mu^{+}}{\sqr{k}{\mathmybb{k}}} \langle k | q | \mathmybb{k} ] + \frac{\mu^{-}}{\agl{k}{\mathmybb{k}}} \langle \mathmybb{k} | q | k ]\right)\ ,
\end{split}
\end{equation}
where $k^\mu=-p_1^\mu-p_1^{\prime \mu} = p_2^\mu + p_2^{\prime \mu}$ and 
\begin{equation}
    q^\mu = \frac{p_1^\mu}{\Delta m_1^2} + \frac{p_2^\mu}{\Delta m_2^2}\ , \qquad k \cdot q = 0\ . 
\end{equation}
Integration over $\theta$ takes the following form and returns   Bessel functions of the first kind:\footnote{Notice that this integral can be used directly to map amplitudes in the $\theta$ basis to amplitudes in the helicity basis.}
\begin{equation}
\label{eq:integral_def_BesselJ}
    \int_{0}^{2\pi} \frac{\dd \theta}{2\pi }\, e^{z e^{i \theta} - \bar{z} e^{-i \theta}} e^{i h \theta} = \frac{(-\bar{z})^{h}}{(z \bar{z})^{h/2}} J_{h}(2 \sqrt{z\bar{z}})\ .
\end{equation}
In particular, recalling $k \cdot q = 0$ (and after performing algebraic manipulations of the spinors) we obtain 
\begin{equation}
    \mathcal{M}(1\,1'\,2\,2') = - \frac{\Lambda^2}{t} J_0 (2 \mu \sqrt{-q^2}) + \mathcal{O}(t)\ .
\end{equation}
\begin{figure}[t]
    \centering
    \includegraphics{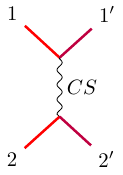}
    \caption{The $t$-channel exchange of a CSP.}
\end{figure}
Restricting to the instructive case $\Delta m_1^2 = \Delta m_2^2 = \Delta m^2$, we have $q^2 = \frac{s}{(\Delta m^2)^2}$ and therefore  
\begin{equation}
 \mathcal{M}(1\,1'\,2\,2')= - \frac{\Lambda^2}{t} J_0 \left(\frac{2 \mu \sqrt{-s}}{\Delta m^2}\right) + \mathcal{O}(t)\ .
\end{equation}

In particular, we may also consider $m_1 = m_2$. 
In this case, we must take into account the Bose symmetry of the scalar and we can provide a class of four-point functions which satisfies the correct factorisation properties:
\begin{equation}
\label{eq:internalcswmass}
    \mathcal{M}(1\,1'\,2\,2') = - \frac{\Lambda^2}{t} J_0\!\left(\frac{2 \mu \sqrt{-s}}{\alpha t + \Delta m^2}\right) - \frac{\Lambda^2}{u} J_0\!\left(\frac{2 \mu \sqrt{-s}}{\alpha u + \Delta m^2}\right)\ ,
\end{equation}
with $\alpha$ a generic parameter. One can even integrate these amplitudes over  $\alpha$ against some measure. A simple example could be:
\begin{equation}
\label{eq:internalcswmass_complex}
    \mathcal{M}(1\,1'\,2\,2') = - \frac{\Lambda^2}{2t} \left[J_0\!\left(\frac{2 \mu \sqrt{-s}}{i t + \Delta m^2}\right) + J_0\!\left(\frac{2 \mu \sqrt{-s}}{-i t + \Delta m^2}\right)\right]+\text{$u$-channel}\ ,
\end{equation}
where we summed over $\alpha=\pm i$.
While these are not the most generic class of functions that satisfy the correct factorisation properties and are well-behaved in the high-energy limit, they provide nevertheless simple examples of amplitudes showing the main features of internal CSP exchange. We highlight three notable examples.
\begin{enumerate}
    \item $\alpha=0$. There is no additional singularity in the Mandelstam variables, but the limit $\Delta m^2\to0$ is ill-defined. We can think of the mass difference as a regulator that cannot be completely removed. In this case, the high-energy limit is good (as the amplitude decays fast enough) but it does not match the $\mu \to 0$ result. 
    \item $\alpha \neq 0$ and $\alpha \in \mathbb{R}$. The Bessel function introduces essential singularities at real values for the Mandelstam invariants: $t,u=-\frac{\Delta m^2}{\alpha}$. This is inconsistent with (analytically extended) unitarity. 
    \item $\alpha \neq 0$ and $\alpha \in i \mathbb{R}$.\footnote{This choice requires to sum over complex conjugate values of $\alpha$ to make the amplitude real analytic.} As shown in equation~\eqref{eq:internalcswmass_complex}, the Bessel function introduces essential singularities in $t,u=-i\frac{\Delta m^2}{\alpha}$ on the imaginary axis. Since analyticity in the upper-half plane is correlated to causality, these amplitudes may be at odd with the latter.  
\end{enumerate}

Notice that the latter two cases admit, at the level of the amplitude, a finite $\Delta m^2\to0$ limit.  Naively, this suggests  a prescription to define amplitudes in the equal-mass case. Nevertheless, these amplitudes exhibit pathological behaviours, as we have discussed and it's further elaborated in Section~\ref{subsec:potential}.

The example discussed in this subsection is instructive at the technical level. 
\begin{enumerate}[label=\alph*.]
    \item As expected, the explicit dependence on the black-board spinors of the intermediate CS disappears (the result cannot depend on the LG phase of the internal particles). In particular, we notice that the kinematic exponents of the in- and out-states combine to give $\frac{\langle k | q | \mathmybb{k} ]}{\sqr{k}{\mathmybb{k}}}$ (and its complex conjugate). It is easy to prove that such a term does not depend on $| \mathmybb{k} ]$, as long as $q\cdot k = 0$ holds. This is a generic feature of sewing internal CSPs. For generic amplitudes, we have always:
	\begin{equation}
		q^\mu = \frac{p_L^\mu}{\langle k | p_L | k ]} - \frac{p_R^\mu}{\langle k | p_R | k ]}\ ,
	\end{equation}
	where the respective minus sign comes from mapping the incoming CSP to an outgoing one ($\theta \to \theta + \pi$), and $L$ and $R$ stand for left and right, respectively.
    \item We pinpoint the appearance of Bessel-$J$ functions as an essential feature of the exchange of a CSP in a channel at tree level.
    \item Finally, we can combine the tree-level amplitudes computed in this subsection (and those in the following) to bootstrap the corresponding one-loop amplitudes from dispersion relations. Checking the conditions for the one-loop amplitudes to be crossing-symmetric, may give constraints on $\alpha$. In particular, it would be interesting to understand whether the loop amplitudes develop IR divergences. Indeed, from the study of three-point amplitudes, we expect it to be free of soft and collinear divergences as the finite mass-splitting provides an IR regulator. IR divergences would put at stake the well-definiteness of perturbative scattering amplitudes for CSPs.
\end{enumerate}

\subsection{The Compton-like (Rayleigh) amplitude}
\label{sec:comptonlikescatt}

An example of a Compton-like amplitude is the absorption and emission of a mostly-scalar CS particle mediated by the excitation of a non-elementary particle with two scalar energy levels $m$ and $m'$.

We start again from the three-point amplitude in equation~\eqref{eq:3pt_scalarsDeltam},
with $\Delta m^2=m'^2-m^2$. The contribution to the 4-point amplitude is:\footnote{In this Section~\ref{sec:comptonlikescatt} we have $s=(p_1+p_2)^2$, $t=(p_1+p_3)^2$, and $(p_2+p_3)^2$. 
}
\begin{equation}
\label{eq:comptolikeEqfiniteM}
    \begin{split}
        \mathcal{M}(1^{\theta_1}2\,3^{\theta_3}4)= - e^{2i \mu\frac{ \epsilon_1 \cdot p_2 + \epsilon_3 \cdot p_4}{s-m^2}} \frac{\Lambda^2}{s-m'^2} - e^{2i \mu\frac{\epsilon_1 \cdot p_4 + \epsilon_3 \cdot p_2}{u-m^2}} \frac{\Lambda^2}{u-m'^2}\ .
    \end{split}
\end{equation}
where for the sake of compactness we expressed the CSPs exponentials in terms of 4-vectors, see equation~\eqref{footnote:epsilonDefinition}. 

\begin{figure}[t]
    \centering
    \includegraphics{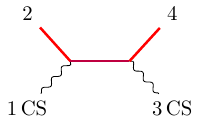}
    \includegraphics{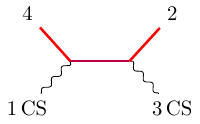}
    \caption{The $s$- and $u$-channel exchanges of excitations mediating a Compoton-like scattering of CSPs.}
\end{figure}

Contrary to the previous examples, this result is {\it uniquely fixed} by matching an ansatz to its factorisation channels. For instance, any deformation of the denominators in the exponentials, \textit{e.g.} $(s-m^2)^{-1}\to [(1+\alpha)s - m^2 - \alpha m^{\prime 2}]^{-1}$, 
and likewise for the $u$-channel, returns identical residues on the factorisation channels but it is not compatible with the general covariant properties under LG transformations, fixed in equation~\eqref{eq:CS_kinematics}. The latter crucially demands the denominator to be $\langle 1 | p_2 | 1] = \langle 3 | p_1 | 3] = s - m^2$ and analogously for the $u-$channel.


\subsection{The partial-wave decomposition of CSPs}
\label{subsec:pw}

In this section, we present the PW decomposition of four (identical) CSPs scattering,\footnote{In principle, we may consider the PW decomposition with other choices of states including non-CS and CS states. The procedure described in the present section adapts easily, just picking the relevant three-point amplitudes from Section~\ref{sec:All3pts}, and including an extra factor of $1/2$ for non-identical particle scattering whenever needed.} following the strategy of reference~\cite{Arkani-Hamed:2017jhn,Jiang:2020rwz}. We decompose the Poincar\'e-reducible 2-CSP state into a sum of irreps of definite angular momentum $j$ at the mass $s=m^2=2p_1\cdot p_2=2p_3\cdot p_4$ ($s$-channel PW-decomposition).  That is, we glue two effective three-point amplitudes \eqref{eq:genera3pt2CSP1MassJ} studied in Section~\ref{sec:All3pts}: 
%
\begin{align}
\label{eq:PWexpansion}
\mathcal{M}(1^{\theta_1}2^{\theta_2} 3^{\theta_3}4^{\theta_4})= & 16\pi\sum_{\{h_i\}} e^{-i(h_1\theta_1+\ldots +h_4\theta_4)} e^{-i(h_{12}+h_{34})\phi} \times  \\
\nonumber
 & \times \sum_{j\geq |h_{12}|\ ,\, |h_{43}|} (2j+1) d^{j}_{h_{12}, h_{43}}(\theta)\mathcal{M}^{(j)}_{h_1 h_2 h_3 h_4}(s) 
\times e^{2i\mu\frac{\epsilon_1\cdot p_2+\epsilon_2\cdot p_1} s}e^{2i\mu\frac{\epsilon_3\cdot p_4+\epsilon_4\cdot p_3}s}\ ,
\end{align}
where $d^j_{h,  h^\prime}(\theta)$ are Wigner $d$-matrices, $\theta$ the scattering angle, $\phi$ the rotation of the scattering plane, and $h_{ij}\equiv h_i-h_j$. The $\mathcal{M}^{(j)}_{h_1 h_2 h_3 h_4}(s)$ are the partial-$j$ amplitudes in the all-incoming convention (the in-out convention reads $\mathcal{M}^{(j)}_{h_1\, h_2\to -h_3\, -h_4}=(-1)^{h_3+h_4}\mathcal{M}^{(j)}_{h_1 h_2 h_3 h_4}$).  
%
%

By direct inspection, the exponential LG-factors on the right-hand side are just equal to 1 in the center of mass (c.o.m.) frame for particle pairs  12 and 34, those with $s>0$
 \begin{equation}
\label{eq:comframeexp}
    e^{2i\mu\frac{\epsilon_1\cdot p_2+\epsilon_2\cdot p_1} s} \overset{\mathrm{c.o.m.}}{=}1 \overset{\mathrm{c.o.m.}}{=} e^{2i\mu\frac{\epsilon_3\cdot p_4+\epsilon_4\cdot p_3}s} \,.
\end{equation}

The PW decomposition  \eqref{eq:PWexpansion}  can be inverted to extract the partial waves. This is done by stripping off from the amplitude the little-group exponential factors that appear in \eqref{eq:PWexpansion}, then Fourier transforming the $\theta_i$, and at this point project with Wigner $d$-matrices,  as in the familiar case, thanks to their orthogonality
\begin{align}
\label{eq:PWexpansionInvertion}
\mathcal{M}^{(j)}_{h_1 h_2 h_3 h_4}(s)=\int_{-1}^1 \frac{\dd \cos\theta}{32\pi} d^{j}_{h_{12}, h_{43}}(\theta)  \int_0^{2\pi} \prod_{i}^4 \frac{\dd \theta_i}{2\pi}e^{i (h_1 \theta_1+\ldots + h_4 \theta_4)}  \frac{\mathcal{M}(1^{\theta_1}2^{\theta_2}3^{\theta_3}4^{\theta_4})\big|_{\phi=0}}{e^{2i\mu\frac{\epsilon_1\cdot p_2+\epsilon_2\cdot p_1} s}e^{2i\mu\frac{\epsilon_3\cdot p_4+\epsilon_4\cdot p_3}s}}\,.
\end{align}

In the c.o.m. frame the LG-exponentials on the right-hand side simplify thanks to \eqref{eq:comframeexp}, and thus the $\theta_i$-Fourier transform convert the integrand to the $h$-basis CSP-scattering amplitudes\footnote{The $\mathcal{M}_{h_1\ldots h_4}$ here should not be mistaken for the ordinary massless amplitudes associated to the $\mu\to0$ limit.} $\mathcal{M}_{h_1 h_2 h_3 h_4}$, see \eqref{eq:hbasisdefinition} and \eqref{eq:changebasistoh}, hence recovering the familiar c.o.m. PW decomposition \cite{Bellazzini:2022wzv}: 
\begin{align}
\mathcal{M}^{(j)}_{h_1 h_2 h_3 h_4}(s)\overset{\text{c.o.m.}}{=}
\int_{-1}^1 \frac{\dd \cos\theta}{32\pi} d^{j}_{h_{12}, h_{43}}(\theta)\mathcal{M}_{h_1 h_2 h_3 h_4}\big|_{\phi=0}\,. 
\end{align} 
Notice that in general the LG-factors associated to the $t$- and $u$-channels are non-trivial and therefore modify the partial-wave expansion w.r.t. the one of the high-energy limit amplitude. 

\subsection{Analytic structure and unitarity}
\label{subsec:analyticityunitarity}

The examples explored so far allow us to infer some general lessons about the structure of four-point amplitudes involving CSPs.

First, by construction, the pole structure is consistent with (extended) unitarity because we used well-defined three-point amplitudes that respect LG scaling and the mass-splitting selection rule. The amplitudes are also crossing-symmetric and satisfy hermitian analyticity. 

Second, the presence of CSPs among initial or final states introduces essential singularities in the Mandelstam variables. Crucially, the latter appear only outside or at the border of the physical region. Their significance w.r.t. extended unitarity deserves further investigation, in connection to the way they enter the PW expansion, see Section~\ref{subsec:pw}.

Third, the exchange of an intermediate --- off-shell --- CSP is connected to the presence of Bessel functions, although the result is less universal as a family of solutions exists; for example, we focus on the one-parameter $\alpha$ family in \eqref{eq:internalcswmass}. In particular, tree-level factorisation is not enough to uniquely fix the argument of the Bessel and, due to the intrinsic non-polynomial structure, EFT arguments can not be invoked. We stress however that for generic choices of the parameter $\alpha\neq0$ that labels the family of solutions for intermediate CSPs, the amplitudes feature extra essential singularities in the Mandelstam variables at generic values in the complex plane. They do not admit an obvious interpretation in terms of extended unitarity. Moreover, when the singularities appear outside the real axis we expect violation of causality.  
In contrast,  a choice without additional singularities in the Mandelstam variables exists, $\alpha=0$ in equation~\eqref{eq:internalcswmass}, but it is  singular in the equal-mass limit. In this case, the mass-splitting acts as a regulator that cannot be removed. An interpretation of finite mass splitting in  terms of off-shell observables is discussed in the next sections.

\section{Weakening the Assumptions}
\label{sec:weakening-Assumptions}

We have seen in previous sections that requiring factorisation in well-defined on-shell three-point amplitudes is heavily constraining,  allowing only a limited set of interactions. For instance, taken at face value, the mass-splitting selection rule together with exact factorisation,  would forbid Compton-like scattering among matter particles of the same mass and CSPs. In this section, we therefore explore whether it is possible --- and what consequences carries--- weakening some of the assumptions. 

One might consider the possibility of starting directly from the four-point amplitudes, according to the general structure given by \eqref{eq:CS_kinematics}. Notice, however, that without the input of factorisation, there are not enough constraints to bootstrap the amplitudes, since the four-point kinematics is less stringent. In particular, there are several allowed and inequivalent choices of the vectors $p^{\pm}_j$ appearing in the exponential LG prefactors, even though different choices correspond to assigning a different overall (kinematically-dependent) phase to the amplitude. In the following, we explore --- with a critical eye but an open mind --- some possible strategies to fix these amplitudes.

\subsection{Massless and degenerate limits}
\label{sec:degenerateLimit}
A first possibility is to consider amplitudes that can be built up using factorisation and then take the massless or equal-mass limit. We explore this possibility in instructive examples.

\subsubsection{Compton-like scattering}
We consider the process described by the amplitude in equation~\eqref{eq:comptolikeEqfiniteM}. Taking now the equal-mass limit $\Delta m^2\to 0$ we get
\begin{equation}
\mathcal{M}(1^{\theta_1}2\,3^{\theta_3}4)= - e^{2i \mu\frac{ \epsilon_1 \cdot p_2 + \epsilon_3 \cdot p_4}{s-m^2}} \frac{\Lambda^2}{s-m^2} - e^{2i \mu\frac{\epsilon_1 \cdot p_4 + \epsilon_3 \cdot p_2}{u-m^2}} \frac{\Lambda^2}{u-m^2}
\end{equation}
This amplitude describes a Compton-like scattering of a mostly-scalar CSP and a massless scalar.

Notice that the essential singularities coming from the exponentials, which before appeared outside the physical region, now overlap with the simple poles at the boundary of the latter. This reflects the absence of the corresponding equal-mass three-point amplitude. The same situation persists if we further take $m\to0$.

The Compton scattering with mostly-photon CSPs  presents peculiarities related to gauge interactions, and it is left to future work.

\begin{figure}[t]
    \centering
    \includegraphics{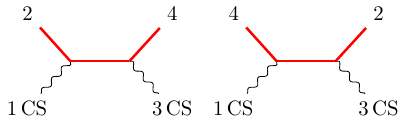}
    \caption{The $s$- and $u$-channel Compoton-like scattering of CSPs in the equal-mass limit.}
\end{figure}

\subsubsection{What about factorisation?}

The previous example highlights a crucial feature of amplitudes with CSPs that \textit{cannot} be obtained via factorisation in three-point amplitudes: the poles that would be related to an on-shell exchange of single particles overlap with the essential singularities that are imposed by little-group covariance. This is directly correlated to the absence of three-point amplitudes violating the mass-splitting selection rule, which in turn implies the absence of simple poles. 
Following backwards the logic of the previous example, we observe that the introduction of a mass splitting regulates the amplitude and allows to disentangle the poles dictated by unitarity from the essential singularities. How can we interpret this limiting procedure?

A possibility could be considering the mass splitting as an off-shellness, and the regulated amplitudes as form factors of fields between physical states. As a concrete example, with reference to the previous section, we may try to interpret the amplitude with non-zero mass splitting as LSZ reduction in a QFT 
\begin{equation}
\begin{split}
    \label{eq:LSZformula}&(2\pi)^{4}\delta^{(4)}\left(p_1+p_2+p_3+p_4\right)\left.\mathcal{M}(1^{\theta_1}2\,3^{\theta_3}4)\right|_{\substack{m_2^2=p_2^2\\m_4^2=p_4^2}}=\\
    &=i(p_2^2-m^2)(p_4^2-m^2)\int \dd^4 x_2\int \dd^4x_4 e^{-ip_2x_2}e^{-ip_4x_4}\bra{0}T\left[\phi(x_2)\phi(x_4)\right]\ket{1^{\theta_1}3^{\theta_3}} \,.
\end{split}
\end{equation}
Indeed, this would imply that taking the limit $p_{2,4}\to m^2$, \textit{i.e.} removing the mass splitting, it returns the amplitude $\mathcal{M}(1^{\theta_1}2\,3^{\theta_3}4)$. Factorisation of the regulated amplitude is then to be interpreted as factorisation of the partially off-shell form factor, which directly descends from unitarity in a QFT, see e.g.~\cite{Weinberg:1995mt}.

The physical picture that emerges from this point of view is the following:  in the presence of CSPs, it is not possible to resolve on-shell intermediate one-particle states as well as external states, \textit{i.e.} the external fields $\phi(x_i)$ in \eqref{eq:LSZformula} cannot be taken on-shell on the factorisation channels. The latter should be considered as sourced at finite distance, \textit{i.e.} as some internal off-shell legs of a larger process. The off-shellness controls how much the rest of the process (\textit{e.g.} the source that produces these particles) can be neglected.\footnote{This is somewhat reminiscent of the IR divergences found in the off-shell approach of reference~\cite{Schuster:2023xqa}; see the discussion around Eq.~(4.9) therein, where the coupling between CSPs and static currents  is actually cut off at times larger than the spin scale.}  Conversely, if we do not probe the exchange of the internal particle on-shell, then we may be able to take the external particles on-shell.\footnote{We notice that this last step can actually be troublesome. Whenever exchanging internal CSPs, the amplitude is not completely fixed and there are choices (namely $\alpha=0$ in the example of Section \ref{sec:CSPexchange}) which become singular when the mass-splitting ---the on-shellness--- is removed.} Nevertheless, full factorisation is recovered in an approximate sense in the high-energy limit. As discussed in Section~\ref{sec:highenergyAndStatistics}, the kinematical LG exponential factors become negligible and the exchange of an on-shell state can be resolved.

This picture is perhaps too na\"ive  since the generalization to more complicated examples, like gauge theories, is highly non-trivial. Moreover, describing CSPs self-interactions requires an off-shell description of the latter, which is lacking at present in our formalism

At the practical level, we can imagine the amplitudes to be regulated by an infinitesimal mass difference, or some amount of off-shellness.  This allows us to disentangle the poles dictated by unitarity from the essential singularities.  It forces us, however, to work in a regime where all the other mass scales or momenta are larger than the regulator, preventing us from lowering them to zero.  For example, this would be an obstruction in taking an exact soft limit. In some cases, one can take a double scaling limit where the momentum is softer than all other scales except for the IR regulator (see the discussion in Section \ref{sec:softlimits}).

In order to explore the effect of essential singularities, within this tentative off-shell prescription,  we consider in the following further examples. In particular  we extract the potential generated by mostly-scalar CSP exchange, and the momentum kick of mostly-photon CSP under gravitational interactions. We also explore the behaviour of these amplitudes in the soft limit.

\subsubsection{Mostly-scalar CS potential}
\label{subsec:potential}

We study the potential generated by the internal exchange of a mostly-scalar CSP. To this purpose we consider the class of amplitudes described in equation~\eqref{eq:internalcswmass}, now keeping $m_1\neq m_2$ to remove the $u$-channel. The amplitude is not completely fixed by factorization, hence we parametrize this freedom via a finite $\alpha\neq0$, as previously discussed, and study some paradigmatic examples where formally $\Delta m^2=0$ hereafter.


\begin{figure}[t]
    \centering
    \includegraphics{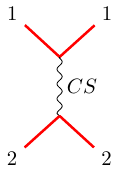}
    \caption{The $t$-channel exchange of a CSP in the equal-mass limit.}
\end{figure}

\par As a general procedure, in order to extract the potential induced by this exchange we evaluate the amplitude in the center-of-mass (CoM) kinematics:
\begin{equation}\label{kinematics1}
	\begin{split}
		&p_1=(\omega_1,\Vec{p}+\frac{\Vec{q}}{2})\ ,\quad p'_1=(\omega_1,\Vec{p}-\frac{\Vec{q}}{2})\ ,\quad p_2=(\omega_2,-\Vec{p}-\frac{\Vec{q}}{2})\ ,\quad p'_2=(\omega_2,-\Vec{p}+\frac{\Vec{q}}{2})\ ,
	\end{split}
\end{equation}
with $\Vec{p}\cdot\Vec{q}=0$, $\omega_{1,2}^2=p^2+\frac{q^2}{4}+m_{1,2}^2$, $s=(\omega_1+\omega_2)^2$ and $t=-\Vec{q}^2$. The potential is then obtained from a three-dimensional Fourier transform of the amplitude in the static limit:
\begin{equation}
	V(\Vec{x})=\frac{-1}{4m_1m_2}\int \frac{\dd^3q}{(2\pi)^3} e^{i \Vec{q}\cdot\Vec{x}}\M(1\,1'\,2\,2')\ .
\end{equation}

\begin{figure}[t]
    \centering
    \includegraphics[]{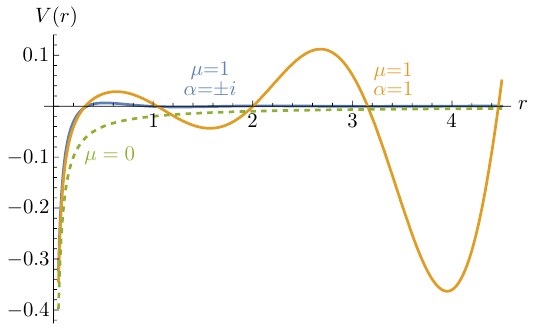}
    \caption{Potential generated by the exchange of a mostly-scalar CSP, with different choices of $\alpha$, compared to the $\mu=0$ case (with $\Lambda=10$, $m_1=m_2=10$).}
    \label{potentialCS}
\end{figure}

We investigate two scenarios, with the essential singularities on the real and the imaginary axis. In particular, we consider the amplitudes with $\alpha = 1$
\begin{equation}
	\label{eq:internalcsalpha}
	\mathcal{M}(1\,1'\,2\,2') = - \frac{\Lambda^2}{t} J_0\!\left(\frac{2 \mu \sqrt{-s}}{t}\right)\ ,
\end{equation}
and with $\alpha = \pm i$
\begin{equation}
    \label{eq:internalcswmass_complex2}
        \mathcal{M}(1\,1'\,2\,2') = - \frac{\Lambda^2}{2t}\left[J_0\!\left(\frac{2 \mu \sqrt{-s}}{i t }\right) + J_0\!\left(\frac{2 \mu \sqrt{-s}}{-i t }\right)\right] \, ,
\end{equation}
where we set $\Delta m^2=0$.
The potential can be analytically computed and expressed in terms of generalized hypergeometric functions, and it is plotted for specific values in Figure~\ref{potentialCS}.   
At short distance $r\mu\ll1$, both cases reduce to the standard scalar potential
\begin{equation}
    V(r)\overset{r \mu \ll 1}{\sim}-\frac{\Lambda^2}{16\pi r m_1m_2}
\end{equation}
In contrast, the behaviour at large distance $r\mu\gg1$ is significantly different depending on $\alpha$. When $\alpha$ is real (and non-vanishing) the potential oscillates with exponentially growing amplitude, anticipating a pathology with this choice, see Fig.~\ref{potentialCS}.  Conversely,  $\alpha$ imaginary gives a potential with polynomially-decreasing oscillatory amplitude. 
Even though constructed in a very different way, we would like to point out that our class of CSP-generated potential is reminiscent of the CSP's potentials found in reference~ \cite{Schuster:2023xqa}.

\subsection{How do CS photons gravitate?}
\label{Sec:Photon-Gravity}
In this section, we would like to address the question of whether we can couple CSPs to standard gravity and compute the \textit{momentum kick} of a CSP off a gravitational field generated by a massive scalar object to leading order in the Newton constant. To this end, we first consider the exchange of a massive spin-2 particle, and then take the massless limit. To single out only the graviton (\textit{i.e.} $h=\pm2$ helicities) in the massless limit, we will consider mostly-photon CSPs in the external states, since they have tracesless energy-momentum matrix elements. 

The relevant three-point amplitudes are
\begin{equation}
    \M(1^{\theta_1} 2^{\theta_2} 3^{j=2})=\frac{\kappa}{2}e^{i\mu\frac{\epsilon_1\cdot p_2+\epsilon_2\cdot p_1}{p_1\cdot p_2}}\cdot 
    \left[
    \frac{\langle 1 \boldsymbol{3}\rangle^4}{\langle 1 2\rangle^2}e^{i(\theta_1-\theta_2)}+\frac{\langle 2 \boldsymbol{3}\rangle^4}{\langle 1 2\rangle^2}e^{-i(\theta_1-\theta_2)}
    \right]\ ,
\end{equation}
and
\begin{equation}
    \M(1 2 3^{j=2})=
    \frac{\kappa}{2}\frac{\langle \boldsymbol{3} | p_1 | \boldsymbol{3}]^2}{M^2}\ ,
\end{equation}
where $M$ is the mass of the graviton, to be sent to zero eventually. The resulting contribution to the four-point amplitude is
\begin{equation}
\label{eq:four_point_gravity}
    \mathcal{M}_4(1^{\theta_1} \phi\, \phi\, 4^{\theta_4})= - \left(\frac{\kappa}{2}\right)^2 e^{2i\mu\frac{\epsilon_1\cdot p_4+\epsilon_4\cdot p_1}{t}} \left[\frac{\langle 1 |p_2|4]^2}{t-M^2} e^{i(\theta_1-\theta_4)} + \frac{\langle 4 |p_2|1]^2}{t-M^2} e^{i(\theta_4-\theta_1)}\right]\ .
\end{equation}
We could have also  obtained the same amplitude, with $M = 0$, simply by dressing the gravitational amplitude with the proper exponential LG factor. In principle, we may consider this as the defining procedure for gravitating  mostly-$h$ CSPs, even for $|h|\neq 1$. This is in fact equivalent to consider off-shell gravitons via the insertion of energy-momentum form factor between two CSPs, see \eqref{eq:Tmunumatrixelement}.

We can now compute the momentum kick to the CSP scattering off the gravitational potential generated by the scalar $\phi$, after taking $M\to 0$. Following the discussion in reference~\cite{Kosower:2018adc}, we consider an initial state which is a superposition of two-particle (the CSP and the scalar) states 
\begin{equation}
    | \psi\rangle_{\rm in} = \int_{0}^{2\pi}\!\dd\theta \int\! \dd\Phi(p_1) \dd\Phi (p_2)\, \Psi(\theta) \psi(p_1) \psi(p_2)\, e^{i p_1\cdot b}\, |p_1 \,\theta,p_2\rangle\ ,
\end{equation}
where $\dd\Phi(p_i)$ is the Lorentz-invariant phase-space measure for the particle $p_i$, $\psi(p_i)$ and $\Psi(\theta)$ are the wavefunctions describing the momenta of the incoming particles and the helicity distribution of the CSP, respectively. $b^\mu$ is the \textit{impact parameter} between the two incoming wavepackets, defined to be spacelike and orthogonal to both wavepacket average momenta. 

\begin{figure}[t]
    \centering
    \includegraphics{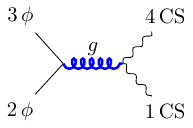}
    \caption{The $t$-channel exchanges of massive graviton.}
    \label{fig:massive_eikonal}
\end{figure}

Then, the momentum kick is given by the difference of the expectation values of the momentum operator in \textit{out} and \textit{in} states:
\begin{equation}
    \Delta p^\mu_{\rm CS} = \, _{\rm in}\langle \psi | S^\dagger [P_1^\mu, S] | \psi\rangle_{\rm in}\ .
\end{equation}
Expanding the $S$-matrix in perturbation theory, at leading order we find\footnote{To leading order in perturbation theory, the scattering angle is computed as $\chi \approx \frac{2 \sqrt{s}}{s-m^2}\frac{\Delta p \cdot b}{\sqrt{-b^2}}$.}
\begin{equation}
\label{eq:KMOCleadingorder}
    \Delta p^\mu_{\rm CS} = \left\langle - i \int\!\frac{\dd^4 q}{(2\pi)^2}\, \delta(2p_1 \cdot q) \delta(2p_2\cdot q) e^{i b\cdot q} q^\mu \left. \mathcal{M}_4(1^{\theta_1} \phi\, \phi\, 4^{\theta_4})\right|_{\substack{p_3 = - p_2 - q\\p_4 = - p_1 + q}}+ \dots \right\rangle\ ,
\end{equation}
where the dots stand for radiative contributions, $q$ measures the \textit{momentum mismatches} of the two wavepackets \cite{Kosower:2018adc}, and $\langle \cdot \rangle$ is a short-hand notation for the integration of the result against the chosen wavepackets in $\theta$.%
\footnote{
\label{footnote:wavepack_average}
We have defined
\begin{equation*}
    \langle f(\theta_1,\theta_4) \rangle = \int_0^{2\pi}\!\dd\theta_1 \int_0^{2\pi}\!\dd \theta_4 \Psi^* (\theta_4) \Psi(\theta_1) \, f(\theta_1, \theta_4)\ ,
\end{equation*}
where $\Psi(\theta)$ is the angular wavepacket of the initial state. In the following, it will be useful to work in the helicity basis and we will also employ the following short-hand notation:
\begin{equation*}
    \langle f(\theta_1,\theta_4) \rangle_{h_1,h_4} = \int_0^{2\pi}\!\frac{\dd\theta_1}{2\pi} \int_0^{2\pi}\!\frac{\dd \theta_4}{2\pi} \, e^{i h_4 \theta_4 - i h_1 \theta_1} \, f(\theta_1, \theta_4)\ .
\end{equation*}
} Indeed, since we are interested in the large-impact-parameter limit, the momentum wavepackets $\psi(p_1)$ have been chosen such that their characteristic lengths are much smaller than the impact parameter. With a slight abuse of notation, the $p_{i=1,2}$ in \eqref{eq:KMOCleadingorder} represent the classical average momenta of the wavepacket, such that both $p_{i}\cdot b=0$.   The large-$b$ limit is equivalent to the limit of large orbital angular momentum, \textit{i.e.} the \textit{eikonal limit}. This will keep only the leading terms in the small-$t$ expansion. Thus, the four-point amplitude must be truncated to the leading order in $t = q^2$.

Before going into performing the Fourier transform, we should consider the wavepackets for the helicity configuration of the CSP in the initial state. It is interesting to focus on two limiting cases, as more general wavepackets can be written as a superposition of those. In particular, we start from the wavefunctions in the helicity basis because it makes the physics more transparent, and then consider the Fourier transform to the $\theta$ basis. We analyze Gaussian and single-helicity distributions:
\begin{equation*}
    \Psi_{\Delta h} (h) = \mathcal{N}_{\Delta h}\, e^{- \frac{(h-\bar{h})^2}{\Delta h^2}}\ , \qquad \Psi_{\hat{h}}(h) = \delta_{h\hat{h}}\ ,
\end{equation*}
where $\mathcal{N}_{\Delta h}$ is a proper normalisation such that $\sum_{h=-\infty}^{+\infty} |\Psi (h)|^2 = 1$. In the $\theta$ basis, the wavepackets become
\begin{equation*}
    \Psi_{\Delta h} (\theta) = \mathcal{N}_\theta\, \Delta h\, e^{-\frac{1}{4} \theta^2 \Delta h^2- i \bar{h} \theta}\, \vartheta_3\!\left(-\pi  \bar{h}+\frac{1}{2} i \pi  \Delta h^2 \theta ,e^{-\pi ^2 \Delta h^2}\right)\ , \qquad \Psi_{\hat{h}}(\theta) = \frac{e^{- i \hat{h} \theta}}{2\pi}\ ,
\end{equation*}
where $\vartheta_3$ is a Jacobi elliptic theta function. For flat helicities distributions $\Delta h \gg 1$, $\Psi_{\Delta h} (\theta)$ becomes sharply peaked around $\theta \sim 0$, \textit{i.e.} we can take $\Psi_{\Delta h} (\theta) = \delta(\theta)$. The single-helicity distributions form an (orthonormal) basis and generic wavepacket can be written as
\begin{equation}
    \Psi(\theta) = \frac{1}{2\pi} \sum_{h=-\infty}^{+\infty} c_h\, e^{- i h \theta}\ , \quad \sum_{h=-\infty}^{+\infty} |c_h|^2 = 1\ .
\end{equation}

The physics in these two cases is qualitatively different. Calculating the expectation value in equation~\eqref{eq:KMOCleadingorder} with a flat distribution is equivalent to taking $\theta_1 = \theta_4$. In this case, we expect the spin-dependent interactions to be averaged away, and the scattering angle recovers twice the GR result for any value of $\sqrt{-b^2} \mu$:
\begin{equation}
    \Delta p^\mu_{\rm CS} \overset{\delta \theta = 0}{=} - \frac{\kappa^2}{16 \pi} (s-m^2) \frac{b^\mu}{b^2} = 2 \Delta p^\mu_{\rm GR}\ .
\end{equation}

\begin{figure}[!t]
    \begin{subfigure}[h]{0.5\linewidth}
    \includegraphics[width=\textwidth]{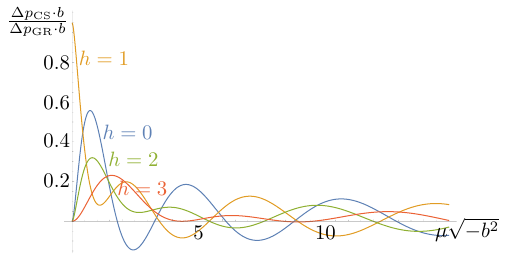}
    \end{subfigure}
    \begin{subfigure}[h]{0.5\linewidth}
    \hspace{-.25cm}
    \includegraphics[]{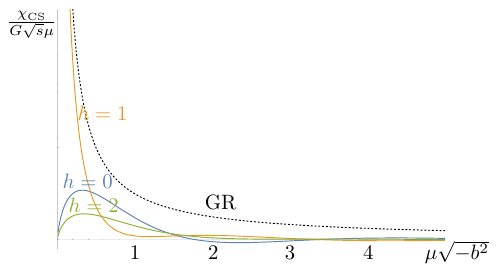}
    \end{subfigure}
    \caption{On the left, the projection of the momentum kick at leading order along the $b^\mu$-direction as a function of $\mu \sqrt{-b^2}$, normalised by the result in GR, for single-helicity wavepackets with $h=0,\dots 3$. On the right, we plot the scattering angle to leading order in the perturbative expansion, normalised with the Schwarzchild radius and $\mu$. For large $\mu \sqrt{-b^2}$, the projection of the momentum kick decays as $\frac{(-1)^h \Delta p_{\rm GR} \cdot b}{2 \pi  \sqrt{\mu \sqrt{-b^2}}} \cos \sqrt{32 \mu \sqrt{-b^2}}$. \label{Fig:scatteringAngle}}
\end{figure}

On the other hand, for a single helicity state $\hat{h}$, we expect to find a qualitatively different result.
\begin{enumerate}
    \item In the UV, only one helicity state is coupled and we have a non-zero deflection iff $\hat{h}=\pm 1$, as expected. The scattering angle is the same as in GR:
        \begin{equation}
        \label{eq:DeltaPUV}
            \Delta p^\mu_{\rm CS} \overset{\sqrt{-b^2} \mu \ll 1}{\sim} 2 \Delta p^\mu_{\rm GR} \left\langle\cos\delta\theta\right\rangle_{\hat{h},\hat{h}} = \begin{cases} \Delta p^\mu_{\rm GR} & \hat{h}=\pm 1 \\ 0 & \text{otherwise} \\ \end{cases}\ .
        \end{equation}
    \item In the IR, all the helicity states are coupled, and we will have a non-trivial contribution from all of them.
\end{enumerate}
Moreover, in the IR, for a generic choice of the helicity wavepacket, the scattering will not stay on the initial scattering plane, determined by $b^\mu$, $p_1^\mu$ and $p_2^\mu$. The momentum kick will have a non-zero component along the orthogonal direction
    \begin{equation}
    \label{eq:v_definition}
        v^\mu = \epsilon^{\mu \nu \rho \sigma} p_{1 \nu} p_{2 \rho} b_{\sigma}\ .
    \end{equation}

We refer to Appendix~\ref{appendix:Fourier} the details of the Fourier transform. Here, we present the final result in terms of the helicity modes of a generic wavepacket:
\begin{equation}
\begin{split}
    \Delta p^\mu_{\rm CS} = \frac{\kappa^2}{64\pi} (s-m^2) \sum_{h_{1},\, h_{4}} c^*_{h_4} c_{h_1} \int_0^{+\infty}\!\dd q
    &\left[J_{h_1-1}\!\left(4\mu/q\right) J_{h_4-1}\!\left(4\mu/q\right)+J_{h_1+1}\!\left(4\mu/q\right) J_{h_4+1}\!\left(4\mu/q\right)\right]\times\ ,\\
    \times (-1)^{\Delta h} \bigg\{  & \left[J_{\Delta h+1}\!\left(\sqrt{-b^2}q\right)-J_{\Delta h-1}\!\left(\sqrt{-b^2}q\right)\right]\frac{b^\mu}{\sqrt{-b^2}} \\
    + i &\left[J_{\Delta h+1}\!\left(\sqrt{-b^2}q\right)+J_{\Delta h-1}\!\left(\sqrt{-b^2}q\right)\right]\frac{v^\mu}{\sqrt{-v^2}}\bigg\}\ .
\end{split}
\end{equation}
The analytic result can be written in terms of Meijer-G functions, but it does not clarify the underlying physics. If we set $\mu = 0$, we recover the result in equation~\eqref{eq:DeltaPUV}, as the first term in the integrand is non-zero iff $h_1=\pm 1$ and $\Delta h=0$. Moreover, if $\Delta h = 0$ --\textit{i.e.} the $\Psi(\theta)$ is a single-helicity wavepacket-- or the $c_h$'s are real, the scattering remains on the initial plane. It is interesting to notice that the scattering angle is non-analytic at $\sqrt{-b^2} \mu = 0$ and, in particular, it has a branch-cut zero to infinity on the negative real axis.

Finally, we can identify the scattering phase with $\delta=-\int_{b}^{\infty} \dd b^\prime \frac{\partial\delta}{\partial b^\prime}=-1/2 \int_{b}^{\infty} \dd  b^\prime \Delta p^\mu \frac{b^\prime_{\mu}}{\sqrt{-b^{\prime\,2}}}$ (with an abuse of notation) for wave-packets that preserve the initial scattering plane, leading to the determination of the time delay $\Delta T=2\partial\delta/\partial\sqrt{s}$. In particular, for definite helicity wave-packets $\Delta h=0$, we find a finite and positive time delay, despite the oscillatory behavior of the scattering angle in Fig.~\ref{Fig:scatteringAngle} which presents alternating regions of attraction and of repulsion. The time delay can be written in terms of Meijer-G functions, reproducing the $-\log(\sqrt{-b^2}\mu)$-scaling of GR for $\sqrt{-b^2}\mu\ll 1$ and $|h|=1$, while approaching positive constants for $|h|\neq 1$. It falls instead polynomially at large scales and arbitrary helicity $h$: 
\begin{equation}
\Delta T_{h}(\sqrt{-b^2}\mu\gg 1)=\left(\frac{\kappa^2 \sqrt{s}}{32\pi^2}\right)\frac{2-(-1)^h\sqrt{2}\sin( 4\sqrt{2}\sqrt{-b^2}\mu)}{\sqrt{-b^2}\mu} > 0\,. 
\end{equation}

We conclude this subsection by emphasizing that CSPs coupled to gravity manifestly violate the \textit{equivalence principle} at large distances, as illustrated in Figure~\ref{Fig:scatteringAngle}, where different helicities fall differently. Remarkably, they do so in a way that remains consistent with General Relativity predictions at short distances. Furthermore, it is surprising that CSPs may even be repelled at long distances without violating causality, as the time delay always remains positive, and even finite.

\subsection{Soft limits}
\label{sec:softlimits}
Universal features of scattering amplitudes typically emerge when external massless particles become soft. In this section, we explore whether this remains true even for CSPs, given the non-trivial competition of two small energy scales: the soft momentum on the one side and the mass splittings, or the off-shellness, on the other side.

\subsubsection{Soft CS radiation}
As a first example, we consider the limit of soft CSPs. For this purpose, we study the following five-point amplitude with three external mostly-scalar CSPs and two massive particles of equal mass:
\begin{equation}\label{eq:5pt1}
    \M(1^{\theta_1}2\,3^{\theta_3}4\,5^{\theta_5})= e^{i \mu\frac{\epsilon_1 \cdot p_2}{p_1\cdot p_2}}e^{i \mu\frac{\epsilon_3 \cdot (p_4+p_5)}{p_3\cdot(p_4+p_5)}}e^{i \mu\frac{\epsilon_5 \cdot p_4}{p_5\cdot p_4}} \frac{\Lambda^2}{(p_1+p_2)^2-m^2}\frac{\Lambda}{(p_4+p_5)^2-m^2}+\text{(permutations)}\ .
\end{equation}
We consider the amplitude in the soft limit $p_5\ll p_i$:
\begin{equation}
    \M(1^{\theta_1}2\,3^{\theta_3}4\,5^{\theta_5})\sim \left[-e^{i \mu\frac{\epsilon_1 \cdot p_2}{p_1\cdot p_2}}e^{i \mu\frac{\epsilon_3 \cdot p_4}{p_3\cdot p_5}}\frac{\Lambda^2}{(p_1+p_2)^2-m^2}\right]\cdot \frac{1}{2 p_4\cdot p_5}\left[-\Lambda e^{i \mu\frac{\epsilon_5 \cdot p_4}{p_5\cdot p_4}}\right]+\text{(permutations)}\ .
\end{equation}
The other orderings are analogous and, as usual, the soft insertions on internal lines are negligible\footnote{The essential singularities do not change this fact, since they are bounded, hence the leading contributions are still determined by the poles.}. Summing all the leading contributions we get
\begin{equation}
    \M(1^{\theta_1}2\,3^{\theta_3}4\,5^{\theta_5})\sim \M(1^{\theta_1}2\,3^{\theta_3}4)\cdot \sum_{i=2,4} \frac{1}{2 p_i\cdot p_5}S_i(5^{\theta_5})\ ,
\end{equation}
where we defined the mostly-scalar CSP soft factor:
\begin{equation}
    S_i(5^{\theta_5})=-\Lambda e^{i \mu\frac{\epsilon_5 \cdot p_i}{p_5\cdot p_i}} \ .
\end{equation}
Thus, the amplitude satisfies factorisation for soft external CSPs. The soft factor has the correct transformation properties under Lorentz and little group and agrees with the results of \cite{Schuster:2013pxj,Schuster:2013vpr}. This conclusion is easily generalized to any amplitude (at least tree-level) built via factorisation and using the prescription for the equal-mass limit:
\begin{equation}
    \M(1...n-1\,n^{\theta_n})\sim \M(1...n-1)\cdot \sum_{i} \frac{1}{2 p_i\cdot p_n}S_i(n^{\theta_n})\ ,
\end{equation}
where index $i$ runs over all the external scalar legs of the hard amplitude.
\par These statements come with a \textit{caveat}: if we are implicitly assuming the amplitude to be regulated by an infinitesimal mass-splitting $\Delta m^2$, then the factorisation in the soft limit should be understood in an approximate sense. Namely, it is valid only in the regime in which the soft momentum is larger than the regulator but much smaller than the other external momenta. Outside this regime, we need to restore the mass-splitting in the formulas above, with the replacement
\begin{equation}
    \frac{1}{2 p_i\cdot p_j}\to\frac{1}{2 p_i\cdot p_j+\Delta m^2} \ .
\end{equation}

Therefore, the enhancement of soft insertions in external legs is not arbitrarily large, but bounded by the $\frac{1}{\Delta m^2}$ factor. As long as $\Delta m^2$ is much smaller than the other (hard) momenta, factorisation still holds. This picture fits with the interpretation of the mass-splitting as an off-shellness: having off-shell external legs means that the amplitude is actually embedded in a larger process in which those legs are actually internal ones. We leave to future work the investigation of beyond-scalar examples.

\begin{figure}[t]
    \centering
    \includegraphics{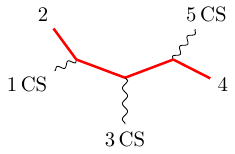}
    \caption{One of the diagrams contributing to the 5-point amplitude of \eqref{eq:5pt1}.}
    \label{5point1}
\end{figure}

\subsubsection{Soft non-CS radiation}
We consider now the case of non-CS soft radiation, in the presence of CSPs.
In particular, we start from the three-point amplitude for two mostly-scalar CSPs and a massive scalar:
\begin{equation}
    \M(1^{\theta_1}2^{\theta_2}3)=\Lambda e^{i \mu\frac{\epsilon_1 \cdot p_2}{p_1\cdot p_2}}e^{i \mu\frac{\epsilon_2 \cdot p_1}{p_2\cdot p_1}}\ .
\end{equation}
We can build the following five-point amplitude via factorisation\footnote{For the gluing of the internal CSP we use the equivalent of the $\alpha=1$ prescription in \eqref{eq:internalcsalpha}.} and then take the massless limit:
\begin{equation}\label{eq:5pt2}
    \begin{split}
        \M(1^{\theta_1}2^{\theta_2}3^{\theta_3}4^{\theta_4}5)
        &=\frac{\Lambda^3}{(p_1+p_2)^2(p_4+p_5)^2}e^{i \mu\frac{\epsilon_1 \cdot p_2}{p_1\cdot p_2}}e^{i \mu\frac{\epsilon_2 \cdot p_1}{p_2\cdot p_1}}e^{i \mu\frac{\epsilon_3 \cdot (p_4+p_5)}{p_3\cdot (p_4+p_5)}}e^{i \mu\frac{\epsilon_{4} \cdot (p_1+p_2+p_3)}{p_4\cdot (p_1+p_2+p_3)}}\cdot\\
        &\cdot J_0\left(2\mu\sqrt{\frac{2 p_3\cdot p_4}{(p_4+p_5)\cdot p_3\,(p_4+p_5)\cdot p_4}}\right)+(\text{permutations})\ .
    \end{split}
\end{equation}

\begin{figure}[t]
    \centering
    \includegraphics{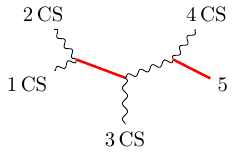}
    \caption{One of the diagrams contributing to the 5-point amplitude of \eqref{eq:5pt2}.}
    \label{5point2}
\end{figure}
In the limit in which particle $5$ is soft, we find
\begin{equation}
    \begin{split}
        \M(1^{\theta_1}2^{\theta_2}3^{\theta_3}4^{\theta_4}5)\sim&
        \frac{\Lambda^3}{(p_1+p_2)^2}e^{i \mu\frac{\epsilon_1 \cdot p_2}{p_1\cdot p_2}}e^{i \mu\frac{\epsilon_2 \cdot p_1}{p_2\cdot p_1}}e^{i \mu\frac{\epsilon_3 \cdot p_4}{p_3\cdot p_4}}\cdot\\
        & \frac{1}{2p_4\cdot p_5}\int_{0}^{2\pi}\frac{d\theta'_4}{2\pi}e^{i \mu\frac{\epsilon'_{4} \cdot p_3}{p_4\cdot p_3}}e^{-i \mu\frac{\epsilon'_{4} \cdot p_5}{p_4\cdot p_5}\frac{p_3\cdot p_4}{p_3\cdot p_5}}e^{i \mu\frac{\epsilon_{4} \cdot p_5}{p_4\cdot p_5}}+(\text{permutations})\ ,
    \end{split}
\end{equation}
where we have expressed the Bessel function in its integral form and $\epsilon'_4=\epsilon_4(\theta'_4)$. Including permutations, we get
\begin{equation}
\M(1^{\theta_1}2^{\theta_2}3^{\theta_3}4^{\theta_4}5)=\sum_{i=1}^{4}\int_{0}^{2\pi}\frac{d\theta'_i}{2\pi}\M_4(...i^{\theta'_i}{j_i}^{\theta_{j_i}}...)\frac{1}{2 p_i\cdot p_5}S_i(5,j_i)\ ,
\end{equation}
with
\begin{equation}
    S_i(5,j_i)=-\Lambda e^{-i \mu\frac{\epsilon'_{i} \cdot p_5}{p_i\cdot p_5}\frac{p_{j_i}\cdot p_i}{p_{j_i}\cdot p_5}}e^{i \mu\frac{\epsilon_{i} \cdot p_5}{p_i\cdot p_5}}\ ,
\end{equation}
where $p_{j_i}$ is the momentum of the CSP that in the hard amplitude enters in the exponential factor of particle $i$ (\textit{e.g.} in our example for $i=4$ we have $j_i=3$.). Hence, the soft factor depends conformally on another momentum of the hard amplitude and loses in this sense its key feature of universality.
As usual, we should think the external scalar to be regulated by a small mass $m$, that can be interpreted as well as an off-shellness. When taking the soft limit, the regulator is simultaneously removed.

\subsection{Backgrounds and pair-wise helicity}
\label{sec:backgrounds}

As we have seen in Section~\ref{sec:bootstappingGeneral}, the main obstruction to building generic three-point amplitudes is the mass-splitting selection rule that emerges from the absence of four-vectors $p^{\pm}_i$ in \eqref{eq:CS_kinematics}, when we restrict to (complex) on-shell kinematics where the three particles' momenta are conserved.  
Nevertheless, a general feature of $p^{\pm}_i$ is that they enter always in a scale-invariant combination, see  \eqref{eq:CS_kinematics}. This suggests that we may look at the on-shell three-point amplitude in the presence of a non-trivial background.

Indeed, in presence of an external four-vector $\xi^{\mu}$ we could easily write a three-point amplitude by setting $p^{\pm}_i=\xi$ for every CS particle:
\begin{equation}
\M_3(\xi)=\prod_{i} \exp\left(\frac{\mu_{i}^{+}}{\sqr{i}{\mathmybb{i}}} \frac{\langle i | \xi | \mathmybb{i} ]}{\langle i | \xi | i ]} + \frac{\mu_{i}^{-}}{\agl{i}{\mathmybb{i}}} \frac{\langle \mathmybb{i} | \xi | i ]}{\langle i | \xi | i ]}\right) \widetilde{\mathcal{M}}_3 \ .
\end{equation}
This background may \textit{e.g.} be a plane wave of momentum $\xi^{\mu}$. If now we try to remove the background sending $\xi\to0$  we notice that the amplitude does not have a well-defined limit, in the sense that it conserves a dependence on the spatial direction.
This is effectively an ordinary four-point amplitude in the soft $\xi$-limit of the fourth (spin-0) leg. 

This ``memory'' of arbitrarily soft momenta suggests that CSPs are quite sensitive to infrared boundary conditions at infinity. For instance, it would be interesting to study CSPs amplitudes in the presence of extended objects, such as \textit{e.g.} strings and monopoles, and possibly uncover a non-trivial interplay between CSPs and topological properties of the environment. We leave this investigation to future work.

In connection with this discussion, we may consider the coupling of CSPs to generic multi-particle states with non-vanishing pair-wise helicity \cite{Csaki:2020yei}. These states can describe \textit{e.g.} monopole scattering and the amplitudes can be described as well using spinor-helicity variables \cite{Csaki:2020inw}. It is unclear whether introducing such states can bring non-trivial kinematics to the three-point amplitudes, and this connection deserves a closer study.

\subsection{Non-analytic amplitudes}
\label{sec:non-analytic}

Up to now we considered the amplitude as an analytic function of the (complex) spinors $|i\rangle$, $|\mathmybb{i}\rangle$, $|i],\ldots$.  The physical value of the amplitude is then recovered as a boundary value. This property is correlated to the notion of \textit{causality}. Nevertheless, from the mathematical point of view, we observe that if we allow the amplitude to depend as well on the complex-conjugate spinors $|i\rangle^*$, $|\mathmybb{i}\rangle^*$, $|i]^*,\ldots$,  we can construct three-point amplitudes consistent with the little-group constraints. In fact we require the weaker condition that the little group scaling is satisfied only under the real section of the complexified Lorentz group. Hence, we impose
\begin{equation}
\label{eq:nonanalyticlg}
    (\alpha_-\mathbb{W}^{+}_i+\alpha_+\mathbb{W}^{-}_i)\mathcal{M}_{\theta_1\ldots \theta_n}=(\alpha_-\mu_i^{+}+\alpha_+\mu_i^{-})\mathcal{M}_{\theta_1\ldots \theta_n}\ ,
\end{equation}
for generic complex $\alpha_-=(\alpha_+)^*$. Indeed, in the presence of complex-conjugate spinors the representation of the complexified Lorentz algebra\footnote{A generic element $T$ of the complexified Lorentz algebra acts on the complexified spinors as
$T\lambda^{*}=(T\lambda)^*$} is no more complex-linear (but only real-linear). We need to restrict to the real Lorentz algebra to obtain a linear representation. In particular, $W_{\mu}^2$ must be expressed as $-W_1^2-W_2^2$.

We consider the example of a three-point amplitude with two standard massless particles (1,2) and one CSP (3). In previous sections, we have seen that analytic solutions to the little-group constraints do not exist. So we start by considering kinematics with $[ij]=0$ and look for non-analytic cases. The only non-trivial brackets under $ISO(2)$ translations are then $\langle\mathmybb{3}1\rangle$ and $\langle\mathmybb{3}2\rangle$, but only a linear combination of the two is independent, thanks to the Schouten identity. Then, consider the following ansatz:
\begin{equation}
\mathcal{M}_{\theta_3}=\exp\left(\frac{\mu_3^{-}}{\langle 3 \mathmybb{3} \rangle} \frac{\langle 1 \mathmybb{3}\rangle }{\langle 1 3\rangle} -\frac{\mu_3^{+}}{\langle 3 \mathmybb{3} \rangle^*} \frac{\langle 1 \mathmybb{3}\rangle^* }{\langle 1 3\rangle^*}\right) \left(\sum_{h_3}\mathcal{M}_{h_3} e^{-ih_3 \theta_3}\right)\ ,
\end{equation}
which satisfies equation~\eqref{eq:nonanalyticlg} and the helicity constraint. It is manifest why we cannot look at the complexified LG algebra: indeed, the amplitude is still annihilated by $\mathbb{W^{+}}$, while the transformation under $\mathbb{W^{-}}$ is such that the real generators $W_1$ and $W_2$ have independent real eigenvalues proportional to $ \mu\cos\theta$ and $\mu\sin\theta$.  Therefore, we see that imposing LG constraints provides us with a prescription to extend the amplitude to complex kinematics, as an alternative to the analytic continuation. This amounts to a different choice of the $i\epsilon$ prescription, corresponding, in our understanding, to the one of \cite{Schuster:2023jgc}. We remark that the non-analyticity is restricted to the exponential factor, hence in the high-energy limit we recover an analytic amplitude. 

We can push this example further and try to build a four-point amplitude. In particular, we consider a mostly-scalar CSP with a coupling to massless scalars. We can take the three-point amplitude in the holomorphic and anti-holomorphic kinematics to be
\begin{equation}
	\begin{split}
		\mathcal{M}^{H}(1\,2\,3^{\theta_3})&=\Lambda\exp\left(\frac{\mu_3^{-}}{\langle 3 \mathmybb{3} \rangle} \frac{\langle 1 \mathmybb{3}\rangle }{\langle 1 3\rangle} -\frac{\mu_3^{+}}{\langle 3 \mathmybb{3} \rangle^*} \frac{\langle 1 \mathmybb{3}\rangle^* }{\langle 1 3\rangle^*}\right)\ ,\\
		\mathcal{M}^{AH}(1\,2\,3^{\theta_3})&=\Lambda\exp\left(\frac{\mu_3^{+}}{[ 3 \mathmybb{3} ]} \frac{[ 1 \mathmybb{3}] }{[ 1 3]} -\frac{\mu_3^{-}}{[ 3 \mathmybb{3} ]^*} \frac{[ 1 \mathmybb{3}]^* }{[ 1 3]^*}\right)\ ,
	\end{split}
\end{equation}
and a natural ansatz for the Compton scattering is
\begin{equation}
\label{eq:non_analytic4pt}
    \begin{split}
        \M(1^{\theta_1}2\,3^{\theta_4}4)=&-\frac{\Lambda^2}{s_{12}}\exp\left(\frac{\mu_1^{-}}{\langle 1 \mathmybb{1} \rangle} \frac{\langle 2 \mathmybb{1}\rangle }{\langle 2 1\rangle} -\frac{\mu_1^{+}}{\langle 1 \mathmybb{1} \rangle^*} \frac{\langle 2 \mathmybb{1}\rangle^* }{\langle 2 1\rangle^*}\right)\exp\left(\frac{\mu_3^{+}}{[ 3 \mathmybb{3} ]} \frac{[ 4 \mathmybb{3}] }{[ 4 3]} -\frac{\mu_3^{-}}{[ 3 \mathmybb{3} ]^*} \frac{[ 4 \mathmybb{3}]^* }{[ 4 3]^*}\right)+\\
        &-\frac{\Lambda^2}{s_{12}}\exp\left(\frac{\mu_1^{+}}{[ 1 \mathmybb{1} ]} \frac{[ 2 \mathmybb{1}] }{[ 2 1]} -\frac{\mu_1^{-}}{[ 1 \mathmybb{1} ]^*} \frac{[ 2 \mathmybb{1}]^* }{[ 2 1]^*}\right)\exp\left(\frac{\mu_3^{-}}{\langle 3 \mathmybb{3} \rangle} \frac{\langle 4 \mathmybb{3}\rangle }{\langle 4 3\rangle} -\frac{\mu_3^{+}}{\langle 3 \mathmybb{3} \rangle^*} \frac{\langle 4 \mathmybb{3}\rangle^* }{\langle 4 3\rangle^*}\right)+(2\leftrightarrow4)\ .
    \end{split}
\end{equation}
A weak form of factorisation can be checked by looking at the limits:\footnote{Since the amplitude is not an analytic function we cannot define properly the residues.}
\begin{equation}
\begin{split}
    &\lim_{\langle12\rangle,[34]\to0}s_{12}\cdot\M(1^{\theta_1}2\,3^{\theta_4}4)\ ,\\
    &\lim_{[12],\langle34\rangle\to0}s_{12}\cdot\M(1^{\theta_1}2\,3^{\theta_4}4)\ ,
\end{split}
\end{equation}
and analogously for $s_{14}$. Taking the first limit we observe that the second line on equation~\eqref{eq:non_analytic4pt} would factorize correctly in the product of three-point amplitudes, nevertheless the first line has essential singularities.\footnote{The exponentials are pure phases by construction, so this amounts to a fastly-oscillating phase.} Taking the second limit the role of the two lines is just reversed. Even if we included only one of the two lines in the amplitude, there would always be a direction in complex kinematics along which the limit is ill-defined.

This example illustrates the difficulty of defining a non-analytic continuation of the scattering amplitude to complex kinematics satisfying even a weak form of factorisation. We leave for future study the investigation of this problem. Furthermore, even when successfully constructing unitary amplitudes, it would be crucial to understand whether they can be consistent with causality, computing an observable such as the time delay.

\section{Discussion and future directions}
\label{sec:discussionConclusions}

In this paper, we initiated the study of continuous-spin particles (CSPs) from the perspective of on-shell consistency conditions. Poincar\'e invariance, little-group covariance, analyticity, and good high-energy behaviour impose stringent constraints, more so than for ordinary massless particles. We found unique solutions for three-point on-shell amplitudes in the space of functions of bi-spinors, reminiscent of the massive ones,\footnote{This is no accident, as the $SU(2)$ little group of massive particles contracts to CSPs $ISO(2)$ for $m\to 0$ and $j\to\infty$ with $mj$ held fixed.} whenever the mass-splitting selection rule is obeyed. In essence, CSPs cannot couple (on-shell, under the given assumptions) to any other pair of particles unless those are non-degenerate in mass. This implies, among other things, that 3-CSPs on-shell amplitudes vanish, as do CSP-particle-antiparticle amplitudes and 2-CSPs-1-graviton amplitudes.

We also classified higher on-shell $n$-point amplitudes and showed that they can be uniquely fixed under favourable conditions. In particular, we bootstrapped on-shell four-point amplitudes for CSPs (with or without ordinary particles) via on-shell factorisation---unitarity---whenever the resulting three-point amplitudes exist. We illustrated these findings through several examples, including the case of intermediate off-shell CSPs. In all cases, the UV amplitudes for CSPs match those of ordinary massless $|h|$-only helicity amplitudes. Infinitely many helicities decouple at short distances, making CSPs a new IR deformation of UV theories, with infinitely many seemingly free degrees of freedom that recouple at low energy.

We also explored various strategies to fix four-point amplitudes when the on-shell three-point function does not exist in a strict sense. Specifically, we examined the consequences of approximate on-shell factorisation, where same-mass particle scattering with CSPs is obtained via a limit of nearly degenerate masses.

Some results obtained this way are however ambiguous. For instance, the potential generated by a CSP exchange between ordinary particles of finite spin or helicity falls into this category, see Section~\ref{subsec:potential}. Removing the mass-splitting may lead to potentials which are not falling with distance, depending on the choice of amplitude. Likewise, the soft emission of CSPs among same-mass particles is only approximate, with the soft factors from \cite{Schuster:2013pxj,Schuster:2013vpr} arising for CSPs momenta softer than any other scales except for the mass splitting of the emitting/absorbing particles.

Conversely, this loose sense of on-shell factorisation works for CSPs exchanging an intermediate graviton. Giving the intermediate graviton a finite mass and eventually removing it (along with all longitudinal polarizations) at the level of physical observables, we obtain well-defined results, such as the momentum-kick (equivalently, the scattering angle) and time delay calculated in Section~\ref{Sec:Photon-Gravity}. This agrees with the fact that the existence of conserved matrix elements $\langle 0| T_{\mu\nu}(0)| 1^{\theta_1}\, \, 2^{\theta_2}\, \rangle$ for the energy-momentum tensor and two CSPs is not inconsistent with the LG scaling, see  \eqref{eq:Tmunumatrixelement}. This, incidentally, provides a loophole in Weinberg-Witten theorem \cite{Weinberg:1980kq} as helicities are no longer Lorentz invariant for CSPs. Clearly, the same construction works for intermediate photon exchange between charged CSPs.

Interestingly, these results show explicitly that CSPs violate the equivalence principle at distances larger than the spin scale, while preserving causality at all scales. In fact, this explains  why the infinitely many helicities of CSPs inside  loops do not renormalize the Newton constant to zero. And it is also the reason why the total cross-sections for producing CSPs remain finite, too.

How do our results align with the intriguing findings of \cite{Schuster:2023jgc, Schuster:2023xqa}? One possibility is that the explicit model presented there---a gauge theory coupled with worldline matter---relaxes some of the basic assumptions about S-matrix theory that we have made in this work. This prompted us to investigate further departures from these assumptions in Section~\ref{sec:non-analytic}, where we relaxed strict analyticity (recovered whenever the Pauli-Lubanski scale vanishes and in the high-energy limit). This would allow keeping the little-group exponential factors carried by CSPs real even for complex kinematics, as done in reference~\cite{Schuster:2023jgc, Schuster:2023xqa} to claim unitarity. However, we have not yet reconciled non-analyticity with on-shell factorisation within the framework of on-shell scattering amplitudes pursued in this manuscript. We believe this important point warrants further analysis, which we leave for future investigations.

Finally, we highlight a couple of directions for further study. While we focused on some consistency conditions, an entire class of stress tests remains, \textit{i.e.} loop corrections. Are cross-sections for CSPs finite or IR divergent? Do we need to discuss inclusive observables rather than exclusive scattering amplitudes? We expect that CSPs are actually better behaved in the IR than ordinary massless particles, see e.g. the finite gravitational time delay found in Section~\ref{Sec:Photon-Gravity}, at least for those interactions that respect the mass-splitting selection rule. 

Another interesting direction, briefly touched upon in Section~\ref{sec:backgrounds}, involves the IR sensitivity of CSPs. Studying CSP theory in the environment of extended objects, such as topological defects, could be insightful. Looking for even more dramatic and speculative IR deformations, giving the CSPs a mass  warrants further study as it would allow to access unambiguously their off-shell dynamics. 

An obvious question is whether one can construct recursive relations, similar to BCFW, to efficiently reconstruct high $n$-point amplitudes from lower ones. This is a non-trivial task due to the essential singularity appearing in the LG factors of CSPs.

Yet another intriguing direction is related to the existence of non-trivial expectations of local correlators evaluated on CSPs states, \textit{e.g.} in \eqref{eq:Jmumatrixelement} and \eqref{eq:Tmunumatrixelement}. It would be interesting to use those matrix elements as seeds of a bootstrap program to extract higher point correlators, such as electromagnetic or gravitational Compton scattering, or  \textit{e.g.} three-point current correlators associated to anomalies. Although the results in Section~\ref{Sec:Photon-Gravity} are promising, preliminary investigations  reveal significant challenges. Additionally, the behaviour of minimally coupled CSP gravitons and photons presents further complexities that warrants future study.

Besides these formal investigations, it would be phenomenologically interesting to understand the thermodynamic  properties of CSPs, see \cite{Schuster:2024wjc} for a first step in this direction. In particular, a natural question is how many effective thermalized degrees of freedom $g_*(T)$ contribute to the entropy density of our universe at any given temperature, given stringent experimental constraints. Although we have firmly established that scattering amplitudes for CSPs make sense, experiments may have already strongly constrained them.

\subsection*{Acknowledgments}
We thank Alfredo Glioti and Pablo Sesma for discussions and early participation. We thank Donal O'Connell for the interesting discussions, and Natalia Toro and Philip Schuster for the comments on the manuscript.  SDA is supported by the European Research Council, under grant ERC–AdG–88541. We also acknowledge the use of AI in improving the grammar and flow of this manuscript. 


\appendix

\section{Conventions, notation, and spinors reminder}
\label{appendix:conventions}
We work in 4-dimensional Minkowski spacetime with  metric $\eta_{\mu\nu}$, signature $(+,-,-,-)$, and fully anti-symmetric  tensor $\epsilon_{\mu\nu\rho\sigma}$ with $\epsilon_{0123}=1$. (Anti-)symmetrization of indices is represented by indices inside round (square) brackets without combinatorical factors, \textit{e.g.} $A^{(a}B^{b)}=A^{a}B^{b}+A^{b}B^{a}$ and $A^{[a}B^{b]}=A^{a}B^{b}-A^{b}B^{a}$. Summation over repeated spinor or Lorentz indices is understood, \textit{e.g.} $\rho^{\alpha}\lambda_{\alpha}$ is short-hand notation for $\sum_{\alpha} \rho^{\alpha}\lambda_{\alpha}$. 

Right-handed and left-handed spinors are vectors in $\mathbb{C}^2$ where the defining representation of   $SL(2,\mathmybb{C})\times SL(2,\mathmybb{C})$ act. Spinors are represented by letters carrying dotted and undotted indices respectively,  so that $\lambda_\alpha$ (left) and $\lambda_{\dot\alpha}$ (right) are two independent spinors despite the abuse of notation in using the same letter. This notation carries over to matrix elements  $(\Lambda_{\alpha}^{\,\,\beta},\Lambda_{\dot\alpha}^{\,\,\dot\beta})\in SL(2,\mathmybb{C})\times SL(2,\mathmybb{C})$, namely $\lambda_\alpha\to \Lambda_{\alpha}^{\,\,\beta}\lambda_\beta$ and $\lambda_{\dot\alpha}\to \Lambda_{\dot\alpha}^{\,\,\dot\beta}\lambda_{\dot\beta}$, where the two matrices are independent. 
Suppressing indices and working in matrix notation we find sometimes convenient to distinguish right-handed spinors by a tilde: $\lambda$ is left-handed while $\widetilde{\lambda}$ is right-handed,  that is transforming  as $\lambda\to \Lambda \lambda$, $\widetilde{\lambda}\to \widetilde{\Lambda}\widetilde{\lambda}$ with $(\Lambda,\widetilde{\Lambda})\in SL(2,\mathbb{C})\times SL(2,\mathbb{C})$.

The connection with the complex Lorentz group comes from the 2-to-1 group isomorphism $(\Lambda,\widetilde{\Lambda})\to \Lambda_{\nu}^{\,\,\,\mu}(\Lambda,\widetilde\Lambda)=\mathrm{Tr}\left(\bar{\sigma}_{\nu} \Lambda \sigma^{\mu} \widetilde{\Lambda}^T\right)/2\in SO(3,1)_{\mathbb{C}}$, where $\sigma^\mu_{\alpha \dot\alpha}=(\mathbb{I},\sigma^i)$ and $\bar{\sigma}^{\mu\, \dot\alpha \alpha}\equiv \sigma^{\mu}_{\beta\dot\beta}\epsilon^{\alpha\beta}\epsilon^{\dot\alpha \dot\beta}=(\mathbb{I},-\sigma^i)$ are built out of the Pauli matrices $\sigma^i$. This follows from $\Lambda\sigma^\mu \widetilde{\Lambda}^T=\sigma^{\nu}\Lambda_{\nu}^{\,\,\,\mu}(\Lambda,\widetilde{\Lambda})$, which is a direct consequence of $\det\left(\sigma^\mu x_\mu\right)=x^2$ being invariant, $\forall x^\mu \in \mathbb{C}^4$, under $\sigma^\mu x_\mu\to \Lambda\sigma^\mu \widetilde{\Lambda}^{T} x_\mu$, since $\det \Lambda=1=\det \widetilde{\Lambda}$, and the fact that $\sigma^\mu$ are a basis of $2\times2$ complex matrices. By Schur's lemma one can prove that only $\pm(\Lambda,\widetilde{\Lambda})$ map to the same Lorentz matrix, so that $SO(3,1)_{\mathmybb{C}}\sim SL(2,\mathmybb{C})\times SL(2,\mathbb{C})/\mathbb{Z}_2$. 


The real double cover of Lorentz group, $SO(3,1)\sim SL(2,\mathmybb{C})/\mathbb{Z}_2$, is obtained by restricting to transformations with $\Lambda_{\dot \alpha}^{\,\,\dot\beta}=\left(\Lambda_{\alpha}^{\,\,\beta}\right)^*$ such that $\lambda_{\dot\alpha}= \pm \left(\lambda_{\alpha}\right)^*$ is a consistent condition under real Lorentz transformations. 

The invariant spinor contractions are defined through the fully anti-symmetric (invariant) tensors $\epsilon^{\alpha\beta}$ and $\epsilon^{\dot{\alpha}\dot{\beta}}$, chosen with matrix elements $\left(\epsilon^{\dot{\alpha}\dot{\beta}}\right)^*=\epsilon^{\alpha\beta}$, and their inverse $\epsilon_{\alpha\beta}$ and $\epsilon^{\dot{\alpha}\dot{\beta}}$, \textit{i.e.} $\epsilon_{\alpha\gamma}\epsilon^{\gamma\beta}=\delta_{\alpha}^\beta$, $\epsilon_{\dot\alpha\dot\gamma}\epsilon^{\dot\gamma\dot\beta}=\delta_{\dot\alpha}^{\dot\beta}$.  In particular, raising and lowering indices defined by  $\lambda^\alpha \equiv \epsilon^{\alpha\beta}\lambda_{\beta}$, $\lambda_\alpha = \epsilon_{\alpha\beta}\lambda^{\beta}$ and $\lambda^{\dot\alpha} \equiv \epsilon^{\dot{\alpha}\dot{\beta}}\lambda_{\dot\beta}$, $\lambda_{\dot\alpha} = \epsilon_{\dot{\alpha}\dot{\beta}}\lambda^{\dot{\beta}}$ lead  to the $SL(2,\mathmybb{C})\times SL(2,\mathmybb{C})$-invariant contractions  $\langle \lambda \rho \rangle=\lambda^\alpha \rho_\alpha=\epsilon^{\alpha\beta}\lambda_\beta \rho_\alpha$ and $[\lambda \rho ]\equiv \lambda_{\dot\alpha} \rho^{\dot\alpha}\equiv \epsilon^{\dot{\alpha}\dot{\beta}}\lambda_{\dot\alpha}\rho_{\dot\beta}=\lambda_{\dot\alpha}\rho^{\dot\alpha}=\epsilon_{\dot\alpha \dot\beta}\lambda^{\dot\beta}\rho^{\dot\alpha}$. 
We also adopt the angle- and square-spinor notation where $|\lambda^i\rangle=\lambda^i_{\alpha}$, $\langle\lambda^i|=\lambda^{i\,\alpha}$,  $|\lambda^i]=\lambda^{i\,\dot\alpha}$ and $[\lambda^i|=\lambda^{i}_{\dot\alpha}$, where $i$ is a labels the particular spinor at hand.  Invariant contractions are thus represented as $[\lambda^i \lambda^j]$ and $\langle \lambda^i \lambda^j\rangle$. 

Let's recall how to go from tensors to spinors and vice versa. A general  irrep $D_{(n,m)}$ of $SL(2,\mathbb{C})\times SL(2,\mathbb{C})$  is  identified by a pair of (non negative) half-integers $(n,m)$ such that $D_{(n,m)}(\Lambda,\widetilde{\Lambda})$ maps 
\begin{equation}
D_{(n,m)}(\Lambda,\widetilde{\Lambda}): \mathcal{O}_{\alpha_1\ldots \alpha_{2n} \dot{\alpha}_1\ldots \dot{\alpha}_{2m}}\to \Lambda_{\alpha_1}^{\,\,\beta_1}\ldots \Lambda_{\alpha_{2n}}^{\,\,\beta_{2n}} \widetilde{\Lambda}_{\dot \alpha_1}^{\,\,\dot\beta_1}\ldots \widetilde{\Lambda}_{\dot{\alpha}_{2m}}^{\,\,\dot{\beta}_{2m}}\mathcal{O}_{\beta_1\ldots \beta_{2n} \dot{\beta}_1\ldots \dot{\beta}_{2m}}
\end{equation}
where $\mathcal{O}_{\alpha_1\ldots \alpha_{2n} \dot{\alpha}_1\ldots \dot{\alpha}_{2m}}$ is fully symmetric in the $\alpha$'s and in the $\dot\alpha$'s, separately. Tensor representations are those for which $n+m$ is integer. 

For instance the $(1/2,1/2)$ irreps acting on $\mathcal{O}_{\alpha \dot{\alpha}}$ is equivalent to the 4-dimensional Lorentz vector representation acting  on $\mathcal{O}^\mu=\bar{\sigma}^{\mu \dot{\alpha}\alpha}\mathcal{O}_{\alpha \dot{\alpha}}/2$. It can be inverted via $\mathcal{O}_{\alpha \dot{\alpha}}=\sigma^\mu_{\alpha\dot{\alpha}}\mathcal{O}_\mu$. That is, acting with Pauli matrices one can convert spinor indices to Lorentz and the other way around as well. This is particularly useful for the 4-momenta of the particles which leads to defining 
\begin{equation}
\sigma^\mu_{\alpha\dot\alpha}p_\mu\equiv p_{\alpha \dot\alpha}\, \qquad p_\mu p^\mu=\det \left(p_{\alpha\dot\alpha}\right)\,. 
\end{equation}
A short-hand notation we adopt in the manuscript is also $\langle \lambda| p |\rho]=\langle \lambda|\sigma^\mu|\rho]p_\mu$.   

Massless particles have $p^2=0$, which allows to define momentum-spinors $\lambda_{\alpha}(p)$ and $\lambda_{\dot\alpha}(p)$ via $p_{\alpha\dot\alpha}=\lambda_\alpha \lambda_{\dot\alpha}$, up to $\lambda_{\alpha}\to w \lambda_{\alpha}$ and $\lambda_{\dot\alpha}\to w^{-1}\lambda_{\dot\alpha}$.  For various massless species labelled by $i$, we adopt the convention where $\lambda^i_{\alpha}=|\lambda^i\rangle=|i\rangle$ and  $\lambda^{i\,\dot\alpha}=|\lambda^i]=|i]$. For example, the contractions are represented as $[ij]$ and $\langle i j\rangle$,  and in particular $\langle i j\rangle [j i]=2p^i\cdot p^j$, which follows from $\mathrm{Tr}\bar{\sigma}^{\mu}\sigma^\nu=2\eta^{\mu\nu}$. 

If the particles are massive, $p^2=m^2\neq 0$ then there are two set of spinors (up to $SU(2)$ little group transformations) $\lambda^{I=1,2}_{\alpha}(p)$ and $\lambda^{\dot\alpha}_{I=1,2}(p)$, associated to each momentum, $p_{\alpha\dot\alpha}=\lambda^I_{\alpha}\lambda_{I\, \dot\alpha}$ represented by {\it bold spinors} \cite{Arkani-Hamed:2017jhn}  $|\boldsymbol{\lambda}\rangle [\boldsymbol{\lambda}|$ or just $|\boldsymbol{i}\rangle [\boldsymbol{i}|$ to differentiate species labelled by $i$. 

In Section~\ref{Sec:differential} we introduced yet another set of spinors: for any massless momentum $p^i_{\alpha\dot\alpha}=\lambda^i_{\alpha}\lambda^i_{\dot\alpha}=|i\rangle [i|$  we assign {\it black-board} spinors (up to $ISO(2)$ transformations \eqref{eq:iso2equivalenceclass}) $\rho^i_\alpha=|\mathmybb{i}\rangle$ and $\rho^i_{\dot\alpha}=[\mathmybb{i}|$ such that the contractions  $\langle i \mathmybb{i}\rangle$ and $[ i \mathmybb{i}]$ are non-vanishing constants, both Lorentz and $SL(2,\mathbb{C})$ invariant.  

Another example of tensor expressed in spinorial basis is given by the Hermitian  2-index anti-symmetric Lorentz generator $J_{\mu\nu}=-J_{\nu\mu}$, which corresponds to the direct sum $(1,0)\oplus (0,1)$ of irreps 
\begin{equation}
J_{\alpha\beta}=\frac{1}{4} J_{\mu\nu}\sigma^{[\mu}_{\alpha\dot\gamma}\sigma^{\nu]}_{\beta\dot\delta}\epsilon^{\dot\gamma \dot\delta}\,,\quad 
J_{\dot\alpha \dot\beta}=\frac{1}{4} J_{\mu\nu}\sigma^{[\mu}_{\gamma\dot\alpha}\sigma^{\nu]}_{\delta\dot\beta}\epsilon^{\delta \gamma}\,,\quad 
J_{\mu\nu}=\frac{1}{4 i}\epsilon_{\mu\nu\rho\sigma}\sigma^{\rho}_{\alpha \dot\beta}\sigma^{\delta}_{\beta\dot\alpha}\left(J^{\alpha\beta}\epsilon^{\dot\alpha\dot\beta}+J^{\dot\alpha\dot\beta}\epsilon^{\alpha\beta}\right)\,.
\end{equation}
From this, it follows that the Pauli-Lubanski vector $W_\mu$ of \eqref{eq:Wdef} can be expressed in the $(1/2,1/2)$ representation $W_{\alpha \dot\alpha}=\sigma^\mu_{\alpha\dot\alpha}W_\mu=\frac{-i}{2}\left(J_{\alpha}^{\,\,\beta} P_{\beta \dot\alpha}- J^{\dot\beta}_{\,\,\dot\alpha}  P_{\alpha \dot\beta}\right)$.

Let's comment also on a couple of important signs. First, the invariant contractions are anti-symmetric, so that $\lambda^{\alpha}\rho_\alpha=-\lambda_{\alpha}\rho^\alpha$ and analogous for the right-handed contractions. 
Therefore,  on the real restriction $\left(\lambda_{\dot\alpha}\right)^*=\lambda_{\alpha}$ we have   $[\lambda\rho]^*=-\langle \lambda\rho \rangle$.   
Second, concerning spinor derivatives, from $\partial\lambda_\alpha/\partial \lambda_\beta=\delta_{\alpha}^\beta$ it follows $\partial\lambda^\alpha/\partial \lambda_\beta=\epsilon^{\alpha\beta}$, and therefore $\partial/\partial\lambda_\alpha=\partial/\partial\lambda^\beta\epsilon^{\beta\alpha}$. This implies that lowering and raising spinor indices of $\lambda_\alpha$ and $\partial/\partial \lambda^\alpha$ differ by a relative minus sign, despite they transform exactly in the same way under $SL(2,\mathbb{C})$. In particular
\begin{equation}
\lambda_\alpha \rho^\alpha=-\lambda^\alpha \rho_\alpha\qquad \mbox{but}\qquad \lambda_\alpha \frac{\partial}{\partial\rho_\alpha}=\lambda^\alpha \frac{\partial}{\partial\rho^\alpha}=|i\rangle \frac{\partial}{\partial |\mathmybb{i}\rangle}=\langle i| \frac{\partial}{\partial \langle \mathmybb{i}|}
\end{equation}
and analogous for dotted indices. 

\subsection*{Outgoing CSPs as negative-energy ingoing CSPs }
Finally, we specify our conventions for treating outgoing states as incoming with negative energy. In order to do so, we want to determine the state $\ket{p',\theta'}$ such that it transforms as $\bra{p,\theta}$ under Poincar\'e transformations. For generic bras and kets, the action of Poincar\'e is given by $\ket{p',\theta'}\to U(\Lambda,a)\ket{p',\theta'}$ and $\bra{p,\theta}\to\bra{p,\theta}U(\Lambda,a)^{\dag}$. First, we consider translations, \textit{i.e.} we choose $U (0,a)=e^{-i P \cdot a}$, which fixes $p^\prime=-p$.  To fix $\theta^\prime$ in terms of $\theta$, we act with a little group transformation $U=e^{i \alpha^{+} \mathbb{W}^{-}(p) +i \alpha^{-} \mathbb{W}^{+}(p)}$. On one side, we get trivially
\begin{equation}
    \bra{p,\theta}\to e^{-i\mu e^{i\theta}\alpha^{-}-i\mu e^{-i\theta}\alpha^{+}}\bra{p,\theta} \,.
\end{equation}
On the other side, we need to fix $\mathbb{W}^{\pm} (-p)$ in terms of $\mathbb{W}^{\pm} (p)$. This depends on the choice spinors transform under $p\to-p$: in general, we can have $\lambda (-p) = e^{i\alpha}\lambda (p)$, $\lat (-p) = e^{i(\pi-\alpha)}\lat (p)$, \textit{i.e.} it is fixed up to a LG transformation.\footnote{We can restrict to the discrete choices $\alpha = 0,\frac{\pi}{2},\pi,\frac{3\pi}{2}$, requiring that we recover the same kinematic variables if we perform this transformation a number of times. For example, if we consider a helicity-fixed vector, $(p,\epsilon^\pm)\to (-p,e^{\pm i (\pi - 2 \alpha)}\epsilon^\pm) \to (p,e^{\mp 4 i \alpha}\epsilon^\pm) = (p,\epsilon^\pm)$. In principle, we should transform the black-board spinors accordingly, \textit{i.e.} $\rho \to e^{- i\alpha} \rho$ and $\rhot \to e^{- i (\pi - \alpha)} \rhot$, but this choice will not play any role in the discussion.} From the explicit expression of $\mathbb{W}^{\pm}(p)$ in equation~\eqref{eq:iso2genbispinorsforapp}, we obtain
\begin{equation}
    \mathbb{W}^{\pm}(-p)=e^{\mp 2i\alpha} \mathbb{W}^{\pm}(p)\ .
\end{equation}
Hence
\begin{equation}
    \ket{-p,\theta'}\to e^{i \alpha^{+} \mathbb{W}^{-}(p) +i \alpha^{-} \mathbb{W}^{+}(p)}\ket{-p,\theta'}=e^{i\mu e^{i(\theta'+2\alpha)} \alpha^{-} +i\mu e^{-i(\theta'+2\alpha)}\alpha^{+}}\ket{-p,\theta'}\,.
\end{equation}
This fixes $\theta'=\theta+\pi-2\alpha$. In particular, for the choice $\alpha=0$ ($\lambda (-p) = \lambda (p)$ and $\lat (-p) = - \lat (p)$) we recover the $\theta'=\theta+\pi$ convention we have used through the paper, while for $\alpha=\frac{\pi}{2}$ ($\lambda (-p) =i \lambda (p)$ and $\lat (-p) = i \lat (p)$) we have $\theta'=\theta$. As a consequence, given the kinematic exponential factor for an ingoing CSP $|k,\theta\rangle$
\begin{equation}
    \exp\left(\frac{\mu e^{-i\theta}}{\langle k\mathmybb{k}\rangle}\frac{\langle\xi\mathmybb{k}\rangle}{\langle \xi k\rangle}+\frac{\mu e^{+i\theta}}{[k\mathmybb{k}]}\frac{[\xi \mathmybb{k}]}{[\xi k]} \right) = \exp\left(i \mu \frac{\epsilon_k \cdot \xi}{k \cdot \xi}\right)\ ,
\end{equation}
the exponent for $\langle k,\theta |$ changes sign irrespectively of choice of $\alpha$:
\begin{equation}
    \exp\left(\frac{\mu e^{-i(\theta+\pi-2\alpha)}}{ e^{2 i \alpha}\langle k\mathmybb{k}\rangle}\frac{\langle\xi\mathmybb{k}\rangle}{\langle \xi k\rangle}+\frac{\mu e^{+i(\theta+\pi-2\alpha)}}{e^{-2 i\alpha}[k\mathmybb{k}]}\frac{[\xi \mathmybb{k}]}{[\xi k]}\right)=\exp\left(-i \mu \frac{\epsilon_k \cdot \xi}{k \cdot \xi}\right)\ .
\end{equation}
In general, one can verify that the amplitude's transformation under in-out exchange is independent of the choice of $\alpha$. 
As a consistency check, we can show that the CS states transform as expected in the helicity basis. Indeed, taking their Fourier transforms, we can show how helicity states transforms by changing from outgoing to incoming:
\begin{equation}
    \langle p,h| \sim \int_{0}^{2\pi}\frac{d\theta}{2\pi}e^{-ih\theta}\ket{-p,\theta+\pi-2\alpha}=e^{ih(\pi-2\alpha)}\ket{-p,-h}\ .
\end{equation}
To conclude, we mention that this procedure fixes the incoming negative-energy states up to an overall phase which is immaterial.

\section{Ordinary massless amplitudes and the redundancy of $\rho$ spinors}
\label{appendix:rhospinors}

\par Let $\A=\mathcal{M}^n_{h_1 \dots  h_C}$ be the stripped amplitude defined in equations \eqref{eq:CS_kinematics}, \eqref{eq:helicity_constraint_Mtilde}, \eqref{eq:generalNpointWtilde}, and $\lambda_{\alpha}$, $\rho_{\alpha}$ the spinors associated to a CS particle,  along with their right-handed counter-parts $
\lambda_{\dot\alpha}$ and $\rho_{\dot\alpha}$, the $\lambda_{\alpha
}\lambda_{\dot \alpha}=p_{\alpha\dot{\alpha}}$ being the null 4-momentum.  
As shown in Section~\ref{subsec:npointAmpl}, the amplitude $\A$ satisfies exactly the same LG constraints imposed on ordinary (non-CSPs) massless amplitudes. In particular, the $\mathmybb{W}^-$ constraint reads
\begin{equation}
    \la_{\alpha}\frac{\partial \A}{\partial \rho_{\alpha}}=0 \,.
\end{equation}
We can always simultaneously diagonalize the $\rho$-dilation operator $\rho_{\alpha}\frac{\partial}{\partial \rho_{\alpha}}$ because it commutes with the Poincar\'e algebra, and focus on its eigenfunctions $\A_k$  
\begin{equation}
    \rho_{\alpha}\frac{\partial \A_k}{\partial \rho_{\alpha}}=k \A_k 
\end{equation}
which span any generic amplitudes $\A=\int dk\, c(k) \A_k$. 
Now, the combination
\begin{equation}
\A^0_k \equiv\langle\la\rho\rangle^{-k}\A_k 
\end{equation}
satisfies 
\begin{equation}
        \la_{\alpha}\frac{\partial \A^0_k}{\partial \rho_{\alpha}}=0\,,\qquad 
\rho_{\alpha}\frac{\partial \A^0_k}{\partial \rho_{\alpha}}=0 \,.
\end{equation}
Since $\lambda_{\alpha}$ and $\rho_{\alpha}$ form a complete basis of the spinor-space, the latter equation implies that $\A^0_k$ does not depend at all on $\rho_{\alpha}$. Consequently, $\A_{k}$ depends on $\rho_{\alpha}$ only through $k$ powers of $\langle\la\rho\rangle$. A similar conclusion holds for right-handed spinors so that $\rho_{\dot\alpha}$ enters only via powers of $[\lambda \rho]$. We have explicitly checked this conclusion for all 4pt amplitudes.
We also recall that these contractions $\langle\la\rho\rangle$ and $[\lambda \rho]$ are just constants set by the choice of normalization. 

\par Incidentally, we notice that the previous argument extends to standard massless particles, whose amplitudes satisfy the same constraints of $\mathcal{M}^n_{h_1 \dots  h_C}$: one does not need to introduce for them the $\rho$ spinors, as they would be redundant. This is nothing but the familiar case of gauge invariance where the amplitudes do not depend on the extra spinor introduced to specify a particular gauge choice.

\section{Details on the Fourier transform in Section~\ref{Sec:Photon-Gravity}}
\label{appendix:Fourier}

It is advantageous to rewrite the Fourier integrals in equation~\eqref{eq:KMOCleadingorder} using the parametrisation introduced in reference~\cite{Cristofoli:2021vyo}:
\begin{equation}
\label{eq:q_decomposition}
    q^\mu = z_1\, \frac{p_1^\mu}{\sqrt{s-m^2}} + z_2\, \frac{p_2^\mu}{\sqrt{s-m^2}} + z_b \frac{b^\mu}{\sqrt{-b^2}} + z_v \frac{v^\mu}{\sqrt{-v^2}}\ ,
\end{equation}
where $v^\mu$ is defined in equation~\eqref{eq:v_definition}, the $z$'s are the new integration variables and $m$ is the mass of the scalar $\phi$. The delta functions fix $z_1=0$ and $z_2=0$. The spinors can be rewritten in terms of Mandelstams:
\begin{equation}
\begin{split}
    \langle 1 |p_2| 4 ]^2 &= - e^{-i \Delta \phi} \left[(s-m^2)^2+s t\right] = - (s-m^2)^2 + \mathcal{O}(q^2)\ , \\
    \langle 4 |p_2| 1 ]^2 &= - e^{+i \Delta \phi} \left[(s-m^2)^2+s t\right] = - (s-m^2)^2 + \mathcal{O}(q^2)\ .
\end{split}
\end{equation}
Thus, we find
\begin{equation}
\label{eq:momentum_kick}
\begin{split}
    \Delta p^\mu_{\rm CS} = \left\langle \frac{i \, \kappa^2}{32 \pi^2
    } \int\!\dd^2 z\, e^{- i \sqrt{-b^2} z_b} e^{- 2i\mu\frac{z_b \frac{(\epsilon_1 -\epsilon_4) \cdot b}{\sqrt{-b^2}} + z_v \frac{(\epsilon_1 -\epsilon_4) \cdot v}{\sqrt{-v^2}}}{z_v^2 + z_b^2}} \left(z_b \frac{b^\mu}{\sqrt{-b^2}}+z_v \frac{v^\mu}{\sqrt{-v^2}}\right) \times\right.\\
    \left. \times \frac{(s-m^2) \cos \delta\theta
    }{z_v^2 + z_b^2 + M^2}\right\rangle \ ,
\end{split}
\end{equation}
where $\delta \theta=\theta_1-\theta_4$ is a measure of the LG-phase mismatch of the two wavefunctions used to compute the expectation value. The polarisations are fixed as in footnote~\ref{footnote:epsilonDefinition}:
\begin{equation}
        \epsilon_4 = \left. \epsilon_1\right|_{\theta_1 \to \theta_1 + \delta \theta} + \mathcal{O} (\sqrt{-q^2})\ ,
\end{equation}
and higher orders $\mathcal{O} (\sqrt{-q^2})$ can be ignored in the leading eikonal approximation.
The integration in the case of a uniform distribution of all the helicity becomes trivial. In fact, $\delta \theta = 0$ implies $\epsilon_4 = \epsilon_1$ and the result recovers the deflection angle in GR as expected:
\begin{equation}
\label{eq:momentum_kick_at_zero_deltatheta}
    \Delta p^\mu_{\rm CS} \overset{\delta \theta = 0}{=} - \frac{\kappa^2}{16 \pi} (s-m^2) \frac{b^\mu}{b^2}\ .
\end{equation}
Moreover, we can expand asymptotically equation~\eqref{eq:momentum_kick} in the short-distance limit and also recover GR at leading order in $\sqrt{-b^2} \mu$:\footnote{When considering the single-helicity state, the momentum kick is reduced to half of the result obtained in equation~\eqref{eq:momentum_kick_at_zero_deltatheta}. This difference stems from the fact that only one of the two helicities contributing to the four-point amplitude~\eqref{eq:four_point_gravity} remains after the integration with the wavepacket and it reproduces the GR computation.}
\begin{equation}
\begin{split}
\label{eq:UV}
    \Delta p^\mu_{\rm CS} \overset{\sqrt{-b^2} \mu \ll 1}{\sim} - \frac{\kappa^2}{16 \pi} (s-m^2) \frac{b^\mu}{b^2}  \left\langle\cos\delta\theta\right\rangle \coloneqq 2 \Delta p^\mu_{\rm GR} \left\langle\cos\delta\theta\right\rangle\ .
\end{split}
\end{equation}
In particular, in the case of a single-helicity distribution, the $\theta_{1,4}$ integrations give a non-zero result iff $\hat{h}=\pm 1$.

For intermediate impact parameters $\sqrt{-b^2} \mu \sim 1$, the momentum is deflected outside the scattering plane, by spin-orbit-like interactions. The large-distance limit is more subtle because it depends on the details of the helicity wavepacket. Thus, we should first perform the integration against the latter and then perform the Fourier transform to impact-parameter space.
The momentum kick for a generic wavepacket is
\begin{equation*}
    \begin{split}
        \Delta p^\mu_{\rm CS} 
        = \sum_{h_{1},\, h_{4}} c^*_{h_4} c_{h_1} \left\langle - i \int\!\frac{\dd^4 q}{(2\pi)^2} \delta(2p_1 \cdot q) \delta(2p_2 \cdot q)\, e^{i b\cdot q}\, q^\mu \mathcal{M}_4(1^{\theta_1} \phi\, \phi\, 4^{\theta_4}) \right\rangle_{h_1,h_4}\ ,
    \end{split}
\end{equation*}
where we adopted the notation introduced in footnote~\ref{footnote:wavepack_average}.
The integration over the $\theta$'s can be performed exactly using the identity~\eqref{eq:integral_def_BesselJ}:
\begin{equation*}
\begin{split}
    \Delta p^\mu_{\rm CS} = \sum_{h_{1},\, h_{4}} c^*_{h_4} c_{h_1} \frac{\kappa^2 (m^2-s)}{64\pi^2} (-i)^{\Delta h+1} \int_0^{+\infty}&\!\dd q \int_0^{2\pi}\!\dd \phi\, e^{- i q \sqrt{-b^2} \sin\phi \sqrt{\epsilon^+\!\cdot u\, \epsilon^-\!\cdot u}} \, \frac{\left[\epsilon^+\!\cdot u\, \epsilon^-\!\cdot u\right]^{\frac{\Delta h + 1}{2}}}{\left[\epsilon^-\!\cdot u\right]^{\Delta h}} u^\mu \times\\
    \times &\left[J_{h_1-1}\!\left(4\mu/q\right) J_{h_4-1}\!\left(4\mu/q\right)+J_{h_1+1}\!\left(4\mu/q\right) J_{h_4+1}\!\left(4\mu/q\right)\right]\ ,
\end{split}
\end{equation*}
where $\Delta h = h_4 - h_1$ and
\begin{equation}
    u^\mu = \sin\phi \frac{b^\mu}{\sqrt{-b^2}}+\cos\phi \frac{v^\mu}{\sqrt{-v^2}}\ .
\end{equation}
The angular integration can be performed analytically if we notice that the polarisations can be decomposed in the same basis as in equation~\eqref{eq:q_decomposition}:
\begin{equation}
    \epsilon^{\pm\, \mu} = \frac{b^\mu}{\sqrt{-b^2}} \mp i \frac{v^\mu}{\sqrt{-v^2}}\ .
\end{equation}
Actually, the LHS is not a Lorentz vector while the RHS is. But, we emphasise that the scalar products $\epsilon^{\pm} \cdot u$ are Lorentz invariant because $p_1 \cdot u = 0$.

\addcontentsline{toc}{section}{References} 

\bibliography{bibs} 
\bibliographystyle{utphys}

\end{document}